\documentclass{cernyrep}
\usepackage{texnames}
\usepackage[T1]{fontenc}
\usepackage{rotating}

\begin{document}

\title{Flavour Physics and CP Violation}

\author{Emi Kou }

\institute{Laboratoire de l'Acc\'el\'erateur Lin\'eaire, Universit\'e Paris-Sud 11, CNRS/IN2P3, Orsay, France}

\maketitle

\begin{abstract}
In these three lectures, I overview the theoretical framework of the flavour physics and CP violation. 
The first lecture is the introduction to the flavour physics. Namely, I give theoretical basics of the weak interaction. I follow also some historical aspect, discovery of the CP violation, phenomenological studies of charged and neutral currents and the success of the GIM mechanism. In the second lecture, I describe the flavour physics and CP violating phenomena in the Standard Model (SM). I also give the latest experimental observation of the CP Violation at the B factories and the LHC and discuss its interpretation. In the third lecture, I  discuss the on-going search of the signals beyond SM in the flavour physics and also  the future prospects. 
\end{abstract}

\def\st{\scriptstyle}
\def\sst{\scriptscriptstyle}
\def\mco{\multicolumn}
\def\epp{\epsilon^{\prime}}
\def\vep{\varepsilon}
\def\ra{\rightarrow}
\def\ppg{\pi^+\pi^-\gamma}
\def\vp{{\bf p}}
\def\ko{K^0}
\def\kb{\bar{K^0}}
\def\al{\alpha}
\def\ab{\bar{\alpha}}
\def\be{\begin{equation}}
\def\ee{\end{equation}}
\def\bea{\begin{eqnarray}}
\def\eea{\end{eqnarray}}
\def\CPbar{\hbox{{\rm CP}\hskip-1.80em{/}}}

\newcommand{\Slash}[1]{\ooalign{\hfil/\hfil\crcr$#1$}}


\section{Introduction}
\label{intro}
The Standard Model (SM) is a very concise model and at the same time a very successful chapter in particle physics. 
In the establishment of the SM, the flavour changing and/or the CP violating phenomena had played a crucial roles. 
On the other hand, there is a very important unsolved question related to the CP violation: {\it how the matter and anti-matter asymmetry of the universe occurs in the evolution of the universe?} Although the Kobayashi-Maskawa mechanism has been successful to explain the CP violation in the flavour phenomena, it is known that the single complex phase introduced in this mechanism is {\it not} enough to solve this problem. Since there is no known way to introduce another source of CP violation in the SM (except for the strong CP phase), we expect that the SM needs to be extended. 
Apart from this issue, there are various reasons to expect physics beyond the SM. The search for a signal beyond SM is a most important task of particle physics today. 

In this lecture, we expose the  theoretical basis of flavour physics in the SM and its phenomenology.  

\section{Weak interaction: fundamentals of flavour physics} 
\label{sec:1} 
\subsection{Quarks and leptons}
The flavour physics concerns the interaction among different fermions, quarks and leptons. Fermions are known to appear in three generations:  \\

\begin{tabular}{|c|c|c|c|}
\hline 
\multicolumn{4}{|c|}{Quarks} \\
\hline 
 & \multicolumn{3}{|c|}{Generation} \\ \cline{2-4}
 \raisebox{1.5ex} {Charge} & I & II & III \\ \hline \hline 
 & $u$& $c$& $t$  \\ 
\raisebox{1.5ex} {+2/3e}  & up & charm & top \\ \hline 
 & $d$& $s$& $b$  \\ 
 \raisebox{1.5ex} {-1/3e} & down & strange & bottom \\ \hline 
\end{tabular}
\ \ \ \ \ \ 
\begin{tabular}{|c|c|c|c|}
\hline 
\multicolumn{4}{|c|}{Leptons} \\
\hline 
 & \multicolumn{3}{|c|}{Generation} \\ \cline{2-4}
 \raisebox{1.5ex} {Charge} & I & II & III \\ \hline \hline 
 & $\nu_e$& $\nu_\mu$& $\nu_\tau$  \\ 
 \raisebox{1.5ex} {0} & \begin{minipage}{1.5cm}{electron \vspace*{-0.15cm}\\neutrino}\end{minipage} &   \begin{minipage}{1.5cm}{muon \vspace*{-0.15cm}\\neutrino}\end{minipage} &   \begin{minipage}{1.5cm}{tau \vspace*{-0.15cm}\\neutrino}\end{minipage} \\ \hline 
 & $e$& $\mu$& $\tau$  \\ 
 \raisebox{1.5ex} {-e} & electron & muon & tau \\ \hline 
\end{tabular} \\ 

As we will see in the following, the interactions between the fermions with difference of charge $\pm 1$ can be described by the {\it charged current} while the interactions between the fermions with the same charge is described by the {\it neutral current}. The examples of such processes are $\beta$ decays, 
$K-\bar{K}$ mixing, $e^+\nu$ scattering process, etc... All these processes are governed by an effective coupling, the so-called Fermi constant $G_F=1.16639(2)\times 10^{-5}$\ GeV$^{-2}$. 
 
\subsection{Charged current}
The history of the weak interaction started from the observation of the continuum spectrum of the 
$\beta$ decay of nucleons in the 1930's: 
\begin{equation}
_ZX \to _{Z\pm 1}X+ e^{\mp}\nu
\end{equation}
where $\nu$ is the neutrino postulated by Pauli. During the next two decades, many new experiments were performed and new particles and new decays were discovered. In particular, the two particles called $\theta$ and $\tau$\footnote{Not to be confused with the $\tau$ lepton!} were quite puzzling. They both contain {\it strangeness} and have very similar properties. Besides, they have different decay patterns: $\theta$ decays into two pions  and $\tau$ into three pions.  For a solution to this problem, Lee and Yang had proposed the {\it parity violation} of the weak interaction that was successfully tested by Wu through the $\beta$ decay of $^{60}Co$. After various experimental tests and theoretical argument, it was suggested that the weak interaction should be of the form of $V-A$ ($V$: Vector current, $A$: Axial vector current). In this theory, the charged current involves  fermions with only left-handed chirality. 
Thus, the weak interaction processes in which charge is exchanged between leptons and leptons/hadrons are well described at low energy by the effective Lagrangian: 
\begin{equation}
\frac{G_F}{\sqrt{2}}J_\mu^{\rm } J^{\mu {\rm }} 
\end{equation}
where $J_\mu^{\rm }=\overline{e}\gamma_\mu (1-\gamma_5)\nu$, $J_\mu^{\rm }=\overline{q_d}\gamma_\mu (1-\gamma_5)q_u$. 
where $q_u$, $q_d$ are the up and down type quarks, respectively. 
One of the problems of this theory at the early time was that the discrepancy in the vector coupling when measuring the decay of radioactive oxygen, $^{14}O$: the coupling constant which was thought to be universal, $G_F$, which is the case for the lepton current, was  0.97$G_F$. In the early 60's, this problem was nicely understood by introducing the so-called Cabibbo angle $\theta_c$: the coupling of $\pi$ and $K$ are different and it is proportional to  $\cos\theta_c$ and $\sin\theta_c$, respectively. Therefore, the hadronic current (with three quarks) is written as: 
\begin{equation}
J_{\mu}^{\rm hadron} = \cos\theta_c\overline{u}_{L}\gamma_{\mu}d_{L} +\sin\theta_c\overline{u}_{L}\gamma_{\mu}s_{L}  
\end{equation}
The measurements of  $^{14}O\to^{14}N+e^+\nu$ and $K\to \pi^0e^+\nu$ lead to a consistent value of the {\it Cabibbo angle}, $\theta_c=0.220\pm 0.003$, which proved the correctness of this expression. 

\subsection{Neutral current}
\begin{figure}
\begin{center}
\resizebox{0.4\columnwidth}{!}{%
  \includegraphics{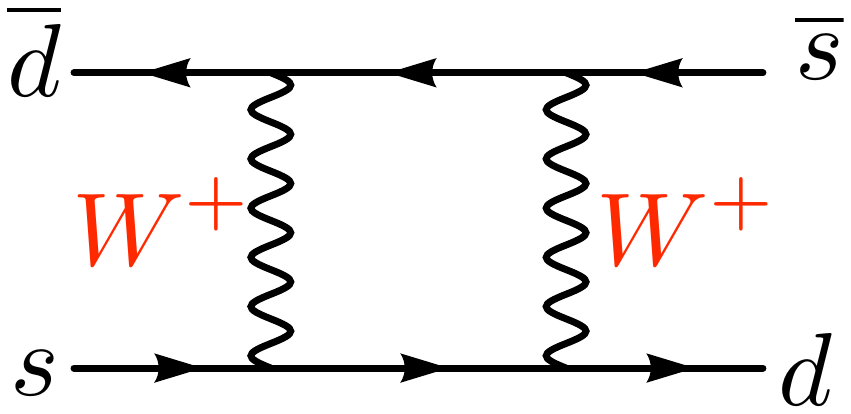} }
\end{center}
\caption{Feynman diagram inducing $K-\overline{K}$ mixing.}
\label{fig:box}       
\end{figure}

From the theory with three quarks, up, down, strange, described above, we can conclude that the quarks provide an $SU(2)$ doublet such as: 
\begin{equation}
Q_L=\Big(\begin{array}{c}u \\ d \cos\theta_c+s\sin\theta_c \end{array}\Big)_L
\end{equation}
Then, the neutral current, namely the term which is induced by $\overline{Q}_L t_3 Q_L$ ($t_3$ is the SU(2) generator) would induce the term proportional to $\overline{d}_L\gamma_\mu s_L$ and $\overline{s}_L\gamma_\mu d_L$, representing strangeness changing neutral current which were not seen in experiments. This problem was solved by Glashow, Iliopoulos and Miani in 1970, by introducing a hypothetical fourth quark, $c$. With this extra quark, one can compose another doublet: 
$\Big(\begin{array}{c}c \\ -d \cos\theta_c+s\sin\theta_c \end{array}\Big)_L$ with which the problematic strangeness changing neutral currents can be cancelled out at the tree level (GIM mechanism). Note however, such flavour changing neutral current can still occur at the loop level if the up quark and the newly introduced charm quark have significantly different masses. Let us see the example of $K-\overline{K}$ mixing.  The diagram is given in Fig.~\ref{fig:box}. This is indeed the strangeness changing ($\Delta S=2$) neutral current.
The amplitude of this process should proportional to: 
\begin{equation}
G_F^2 \Big[(\sin\theta_c \cos\theta_c )^2 f(m_u)-2(\sin\theta_c \cos\theta_c )^2 f(m_u,m_c)+(\sin\theta_c \cos\theta_c )^2 f(m_c)\Big]
\end{equation}
where the first and third terms represent the diagram with either $u$ or $c$ quark in the loop, respectively, while the second term is with both $u$ and $c$ quarks in the loop. The function $f$ is called {\it loop function}, which contains the result of this loop diagram computation and is a function of the internal particle masses (quarks and $W$ boson in this case). If the mass of the up and charm quarks are the same, the three loop functions in this formula coincide, thus, the full amplitude becomes zero (GIM mechanism at one loop level). 
In reality, the observed difference in the up and charm quark masses are significantly different, which can yield non-zero $K-\overline{K}$ mixing. What is remarkable about the work by GIM  is that the fourth quark, $c$, was predicted in order to solve the problem of $K$ decays. It took a couple of years since then but indeed the $c\bar{c}$ charm bound state, $J/\psi $ was discovered in 1974. 

\subsection{Describing the weak interactions in the SM}
The $V-A$ theory developed to explain the $\beta$ decay and strangeness changing interactions is neither renormalizable field theory nor gauge theory. The heavy vector particles which can intermediate the weak interactions is now known as $W$ boson. 
In the late 60's, the model which unifies the electromagnetic interaction and the weak interaction  were developed by S. Glashow, A. Salam and S. Weinberg.  In this model, the $W$, $Z$ and $\gamma$ can be understood as the gauge bosons of the $SU(2)_L\times U(1)_Y$ gauge group. 

In the SM, the masses of the particles are  obtained through the Higgs mechanism, where the $SU(2)_L\times U(1)_Y$ symmetry breaks spontaneously to $U(1)_{\rm EM}$ (while keeping the photon massless). 
Let us see the term which gives the masses of the quarks in the SM, the so-called Yukawa interaction term: 
\begin{equation}
{\mathcal{L}_Y}=\sum_{ij}Y_{ij}^u\overline{\Big(\begin{array}{c}U_i \\ D_i\end{array}\Big)_L} \Big(\begin{array}{c}\phi^0 \\ -\phi^-\end{array}\Big)u_{jR} 
+\sum_{ij}Y_{ij}^d\overline{\Big(\begin{array}{c}U_i \\ D_i\end{array}\Big)_L} \Big(\begin{array}{c}-\phi^{+} \\ \phi^{0}\end{array}\Big)d_{jR}+h.c. 
\end{equation}
which is invariant under $SU(3)_C\times SU(2)_L\times U(1)_Y$ gauge transformations. 
The indices $i,j=1,2,3$ run through the generation. 
The so-called Yukawa matrix is a completely general complex matrix and not constrained in any way (it could be even non-Hermitian).  
Then, after the neutral part of the Higgs  field acquires the vacuum expectation value, the quark mass matrices  are produced: 
\begin{equation}
{\mathcal{L}_Y}=\sum_{ij} m_{ij}^u\overline{U_{iL}}  u_{jR} 
+\sum_{ij} m_{ij}^d\overline{D_{i_L}} d_{jR}+h.c. 
\end{equation}
where 
\begin{equation}
m_{ij}^u=Y_{ij}^u \langle \phi^0\rangle_{\rm vac}, \quad 
m_{ij}^d=Y_{ij}^d \langle \phi^0\rangle^*_{\rm vac}
\end{equation}
This Yukawa mass term can induce parity- and flavour-non-conserving terms. However, we can introduce new quark fields 
\begin{equation}
U^{\prime}_L=K_L^UU_L, \quad u^{\prime}_R=K_R^Uu_R, \quad D^{\prime}_L=K_L^DD_L, \quad d^{\prime}_R=K_R^Dd_R 
\end{equation}
where the matrices $K$  are constrained only by the condition that they must be unitary in order to preserve the form of the kinetic term. Then, when we re-write the mass term with the prime fields, it takes the same form as above but the new matrix: 
\begin{equation}
m^{U\prime} =K_L^Um^UK_R^{U\dagger}, \quad m^{D\prime} =K_L^Dm^DK_R^{D\dagger}
\end{equation}
Now it is a general theorem that for any matrix $m$, it is always possible to choose unitary matrices $A$ and $B$ such that $AmB$ is real and diagonal. Note here that if the matrix $m$ was Hermitian, we would find $A=B$. Therefore, we choose $m^{U\prime}$ and  $m^{D\prime}$ being real and diagonal. Then, the Yukawa mass term does no longer produce the flavour-non-conserving terms, while now the charged current would require some modifications. Let us write the weak doublets with the new prime fields: 
\begin{equation}
Q_{iL}=\Big(\begin{array}{c}(K_L^{U-1}U_L^{\prime})_i \\(K_L^{D-1}D_L^{\prime})_i\end{array}\Big)
\end{equation}
The, the charged current reads: 
\begin{equation}
J^{\mu +}=\sum_{i,j}\frac{1}{\sqrt{2}}\overline{U_L^i}\gamma^{\mu}D_L^j=
\sum_{i,j}\frac{1}{\sqrt{2}}\overline{U_L^{\prime i}}\gamma^{\mu}(K_L^{U\dagger} K_L^D)_{ij}D_L^{\prime j}=
\sum_{i,j}\frac{1}{\sqrt{2}}
\overline{U_L^{\prime i}}\gamma^{\mu}V_{ij}D_L^{\prime j}
\end{equation}
where the unitary matrix $V_{ij}\equiv (K_L^{U\dagger} K_L^D)_{ij}$ is known as Cabibbo-Kobayashi-Maskawa matrix. The rotation of the 1-2 part of this matrix corresponds to the Cabibbo angle discussed above. Now it is clear that the quark mixing which differentiates the $G_F$ to $0.97G_F$ in the hadronic $\beta$ decay originated from the mismatch between the weak eigenstate and mass eigenstate in the SM. 

The full Lagrangian for the quark coupling to the gauge bosons reads\footnote{Here we show only the result for the first generation but the remaining parts can be derived easily by repeating it with different generations.}: 
\begin{eqnarray}
\mathcal{L}&=& \sum_{i}\left[\overline{E_L}(i\Slash{\partial})E_L +\overline{l_{iR}}(i\Slash{\partial})l_{iR} + \overline{Q_{iL}}(i\Slash{\partial})Q_{iL}  
+\overline{u_{iR}}(i\Slash{\partial})u_{iR}+ \overline{d_{iR}}(i\Slash{\partial})d_{iR}\right. \nonumber \\
&& \left. +g (W_{\mu}^+J^{\mu +}_W+W_{\mu}^-J^{\mu -}_W+Z_{\mu}^0J^{\mu}_Z) +eA_{\mu} J^{\mu}_{\rm EM}\right] 
\end{eqnarray}
where the coupling $g$ is related to the Fermi constant by $G_F=\frac{g^2}{4\sqrt{2}M_W^2}$. 
The index $i=1,2,3$ is the generation number. 
The left handed fermions compose $SU(2)$ doublet as: 
\begin{equation}
E_{iL}=\Big(\begin{array}{c}\nu_{i} \\ L_{i}\end{array}\Big)_L, \quad 
Q_{iL}=\Big(\begin{array}{c}U_i \\ D_i \end{array}\Big)_L
\end{equation}
Note that the assignment of the hypercharge $Y$ is $Y=-1/2$ for $E_{iL}$ and $Y=+1/6$ for $Q_{iL}$, which together with $T^3=\pm 1/2$, gives a correct charge $Q=T^3+Y$. For the right-handed fields, $T^3=0$ and thus the hypercharge is equal to the electric charge. Then, the charged, neutral and electro-magnetic currents are written as: 
\begin{eqnarray}
J_{W}^{\mu +} &=& \frac{1}{\sqrt{2}} (\overline{\nu_{iL}}\gamma^{\mu}
L_{iL}+\overline{U_{iL}}\gamma^{\mu}D_{iL}) \\
J_{W}^{\mu -} &=& \frac{1}{\sqrt{2}} (\overline{L_{iL}}\gamma^{\mu}\nu_{iL}+\overline{D_{iL}}\gamma^{\mu}U_{iL}) \\
J_{Z}^{\mu} &=& \frac{1}{\sqrt{\cos\theta_w}}\left[ \overline{\nu_{iL}}\gamma^{\mu}(\frac{1}{2})\nu_{iL}+\overline{L_{iL}}\gamma^{\mu}(-\frac{1}{2}+\sin^2\theta_w)L_{iL}+\overline{l_{iR}}\gamma^{\mu}(\sin^2\theta_w)l_{iR} \right. \nonumber \\
&& +\overline{U_{iL}}\gamma^{\mu}(\frac{1}{2}-\frac{2}{3}\sin^2\theta_w)U_{iL}+\overline{u_{iR}}\gamma^{\mu}(-\frac{2}{3}\sin^2\theta_w)u_{iR} \nonumber \\
&& +\overline{D_{iL}}\gamma^{\mu}(-\frac{1}{2}+\frac{1}{3}\sin^2\theta_w)D_{iL}+\overline{d_{iR}}\gamma^{\mu}(\frac{1}{3}\sin^2\theta_w)d_{iR} \left.\right]\\
J_{\rm EM}^\mu &=& \overline{L_i}\gamma^{\mu}(-1)L_i+\overline{U_{i}}\gamma^{\mu}(+\frac{2}{3})U_i+\overline{D_i}\gamma^{\mu}(-\frac{1}{3})D_i 
\end{eqnarray}
The {\it weak angle} $\theta_w$ relates different couplings and masses, e.g. $g=\frac{e}{\sin\theta_w}$ and $m_W=m_Z\cos\theta_w$. 

\section{CP violation }
\subsection{Matter anti-matter asymmetry in nature}
Back in 1920's, having the theory of relativity of Einstein, Dirac extended the quantum mechanics to incorporate the matter which moves with close to the speed of light. The relativistic quantum mechanics follows the equation of motion called Dirac equation. This equation had one solution which correspond to the electron and in addition, another one that has the same mass and spin as the electron but with opposite charge, an anti-particle. A couple of years after, in 1932, Anderson discovered a particle in  cosmic rays, which indeed corresponds to this solution, a positron! In Dirac's theory, anti-particles and particles can be created and annihilated by pairs. Then, a serious question raises: {\it why only particles (electron, proton, etc) can exist in the universe but not anti-particles? } This theoretical problem has not been solved yet. It seems that something has happened in the early universe, which caused an unbalance between particles and anti-particles. 

Our universe was born about $135\times 10^{11}$  years ago, with extremely high temperature, $10^{19}$ GeV (about 4000 K). After its birth, the universe started expanding. As a result, the temperature dropped rapidly.  At the early time when the temperature was high, the high energy photon could pair-create particles and anti-particles (namely, proton/anti-proton, neutron/anti-neutron, electron/anti-electron). 
At the same time, since all the particles are relativistic, they could also pair-annihilate. As a result, the photon, particle, anti-particle are created and annihilated freely (equilibrium state). Once the temperature reached about  1 MeV, the photon energy was not high enough to create the (anti-)particles. Then, only pair annihilation would have occurred and our universe would not have had any (anti-)particles! However, that has not been the case. For some reasons, by that time, there existed some more particle than anti-particles. The remaining particles composed Helium and then, various nucleus were generated through nuclear interactions. So far, the reason of the asymmetry of number of particle and anti-particle is not known. The only thing we know is that there was some cause of asymmetry when the temperature of the universe was about $10^{15}$ GeV. And in order for this to  happen, there are three conditions (Sakharov's conditions): i) Baryon number  violation, ii) {\it C-symmetry and CP-symmetry violation}, iii) Interactions out of thermal equilibrium. 

It turned out that CP symmetry is violated in nature. This is the subject of this section. The observed CP violation in nature is explained well in the framework of the SM. However, it has also been found that the source of CP violation in SM is much too small to explain the matter anti-matter asymmetry of the universe. This is one of the reasons why we strongly believe that there is physics beyond the SM and why we search for further CP violating observables. \\

\subsection{CP violation in the kaon system}
The first observation of CP  violation was through the measurement of kaon decays. The kaon decays had unusual properties such as the $\theta-\tau$ puzzle as mentioned earlier. 
The kaons came as two isodoublets $(K^+, K^0)$ and their anti-particles $(K^-, \overline{K}^0)$ with strangeness $+1$ and $-1$. The difficulty of assigning the $\theta$ and the $\tau$ to one of $K^0$ or $\overline{K}^0$ is that $\theta$ which decays to two pions should be CP even and $\tau$ which decays to three pions should be CP odd while $K^0$ and $\overline{K}^0$ are both CP even\footnote{Remember: ${\mathcal{CP}}|K^0\rangle=|\overline{K}^0\rangle $, ${\mathcal{CP}}|\overline{K}^0\rangle=|{K}^0\rangle$, ${\mathcal{CP}}|\pi^0\rangle=-|\pi^0\rangle $, ${\mathcal{CP}}|\pi^+\pi^-\rangle=+|\pi^+\pi^-\rangle $,  ${\mathcal{CP}}|(\pi^+\pi^-)_l\pi^0\rangle=(-1)^{l+1}|(\pi^+\pi^-)_l\pi^0\rangle $ where $l$ is angular momentum between $\pi^+\pi^-$ system and $\pi^0$.}. In 1955, it was proposed by Gell-Mann and Pais that the observed state must be the linear combination of the $K^0$ and $\overline{K}^0$ such as: 
\begin{equation}
K_1=\frac{1}{\sqrt{2}} \left(K^0+\overline{K}^0 \right), \quad 
K_2=\frac{1}{\sqrt{2}} \left(K^0-\overline{K}^0 \right). \label{eq:kaonC}
\end{equation}
From the weak interaction point of view, this is quite natural since the weak interaction does not distinguish the strangeness, $K^0$ and $\overline{K}^0$ always mix. Now the problem is solved:  $K_1$ is indeed CP even and $K_2$ is CP odd\footnote{This choice of the kaon state was originally proposed 
based on the idea that ${\mathcal{C}}$ is conserved in weak interaction (notice $K_1$ and $K_2$ are ${\mathcal{C}}$ eigenstates as well). However, when the parity violation in the weak interaction was suggested by Lee and Yang, it was also suggested that charge invariance is also broken, although it was thought that ${\mathcal{CP}}$ was still a good symmetry in weak interaction.  }, which therefore can correspond to $\theta$ and $\tau$, respectively. It is important to notice here that the life time of $K_1$ and $K_2$ are very different. The masses of $K$ being 498 MeV and $\pi$ 140 MeV, the three pion final state is suppressed by the small phase (about a factor 600). This reflects in to the lifetime of these particles: $\tau(K_1)\simeq 0.90\times 10^{-10}\ s$ and $\tau(K_2)\simeq 5.1 \times 10^{-8}\ s$. 
This {\it accidental phase space suppression} will play a crucial role for discovering the CP violation in kaon system. 

In 1962, the experiment of Cronin, Fitch and his collaborators announced the very surprising result that  the long-lived kaon, i.e. $K_2$, decays into two pions: 
\[K_2 \to \pi^+\pi^- \]
Since $K_2$ is CP odd state while two pion is the CP even state, CP is not conserved in this process! 
The fraction is rather small, $2\times 10^{-3}$ of total charged decay modes. Nevertheless, this is the proof that CP invariance is violated in nature! 

A modification to Eq. (\ref{eq:kaonC}) is in order. Now, we name the short- and long-lived kaons as $K_S$ and $K_L$, then, 
\begin{eqnarray}
K_S&=&\frac{1}{\sqrt{2}} \left(pK^0+q\overline{K}^0 \right)
=  \frac{p}{2}\left[(1+\frac{q}{p})K_1+(1-\frac{q}{p})K_2\right] \\
K_L&=&\frac{1}{\sqrt{2}} \left(pK^0-q\overline{K}^0 \right)
=  \frac{p}{2}\left[(1-\frac{q}{p})K_1+(1+\frac{q}{p})K_2\right]
\end{eqnarray}
CP violation ($K_{S,L}\neq K_{1,2}$) occurs when $q/p\neq 1$.

\subsection{Mixing the two kaon states}
Let us now formulate these two kaon states in quantum mechanics. The mixing of the two states comes  from the weak interaction which changes the flavour. Let us describe the time evolution of the $K\overline{K}$ system in terms of the Hilbert space $|\Psi(t)\rangle = a(t)|K\rangle +b(t)|\overline{K}\rangle$ (here we ignore the multi-particle states). The time dependence of this oscillation can be described by the Schr\"{o}dinger equation as: 
\begin{equation}
i\Slash{h}\frac{\partial}{\partial t}\Psi (t)={\mathcal{H}}\Psi(t). 
\end{equation}
where 
\begin{equation}
\Psi(t)=\left(\begin{array}{c}a(t) \\ b(t)\end{array}\right)
\end{equation}
The matrix ${\mathcal{H}}$ is written by 
\begin{equation}
{\mathcal{H}}={\bf M}-\frac{i}{2}{\bf \Gamma}=\left(\begin{array}{cc}M_{11}-\frac{i}{2}\Gamma_{11} & M_{12}-\frac{i}{2}\Gamma_{12} \\
M_{21}-\frac{i}{2}\Gamma_{21} & M_{22}-\frac{i}{2}\Gamma_{22}  \end{array}\right)
\end{equation}
The CPT or CP invariance imposes $M_{11}=M_{22}, \Gamma_{11}=\Gamma_{22}$ and CP or T invariance imposes $\Im\ M_{12}=0=\Im\ \Gamma_{12}$. Then, the eigenvalues and the eigenvectors of this matrix read: 
\begin{eqnarray}
{\rm System\ 1: \ }&
M_{11}-\frac{i}{2}\Gamma_{11}+\frac{q}{p}\left(M_{12}-\frac{i}{2}\Gamma_{12}\right), \quad 
 \left(\begin{array}{c}p \\ q\end{array}\right) & \\
{\rm System\ 2: \ } & 
M_{11}-\frac{i}{2}\Gamma_{11}-\frac{q}{p}\left(M_{12}-\frac{i}{2}\Gamma_{12}\right), \quad 
\left(\begin{array}{c}p \\ -q\end{array}\right)  & 
\end{eqnarray}
which leads to 
\begin{eqnarray}
|K_1\rangle &=& p |K\rangle +q|\overline{K}\rangle \\ 
|K_2\rangle &=& p |K\rangle -q|\overline{K}\rangle  
\end{eqnarray}
with 
\begin{equation}
\frac{q}{p}=\pm\sqrt{\frac{M_{12}^*-\frac{i}{2}\Gamma_{12}^*}{M_{12}-\frac{i}{2}\Gamma_{12}}}
\end{equation}
where the choice of the solution to this equation, either $+$ or $-$, corresponds to replacing the Systems 1 and 2.  Now, it became clear that the CP violation $q/p\neq \pm 1$ occurs when $M_{12}$ and/or $\Gamma_{12}$ is complex number.

The masses and the widths yield: 
\begin{eqnarray} 
& M_1-\frac{i}{2}\Gamma_1\equiv M_{11}-\frac{i}{2}\Gamma_{11}+\frac{q}{p}\left(M_{12}-\frac{i}{2}\Gamma_{12}\right) & \\
& M_2-\frac{i}{2}\Gamma_2\equiv M_{11}-\frac{i}{2}\Gamma_{11}-\frac{q}{p}\left(M_{12}-\frac{i}{2}\Gamma_{12}\right) & 
\end{eqnarray} 
where $M_{1,2}$ and $\Gamma_{1,2}$ are real numbers. 
Here we choose the $+$ sign for the solution for $q/p$ above, and then we define the mass and the width differences as: 
\begin{equation}
\Delta M\equiv M_2-M_1, \quad \Delta \Gamma= \Gamma_1-\Gamma_2
\end{equation}
These two quantities are very important observables for the mixing system. Note that the discussions are totally general and can apply to $D\overline{D}$ and $B\overline{B}$ systems. 

\subsection{Time evolution master formula }
Now let us describe the time evolution of the kaons decaying into pions. 
When there is a mixing of two states, these two states oscillate as time evolves. The CP violating phenomena observed in the kaon system implies that the oscillation rate is different for the state which was $K$ at a given time from those with $\overline{K}$. There is another possibility: the CP violation occurs in  the decays, i.e. the decay rates of $K$ and $\overline{K}$ are different. To summarize, there are a two possibilities of source of the CP violation: 
\begin{equation}
{\rm Oscillation:\ \ \ }K\stackrel{\Slash{\rm CP}}{\longleftrightarrow}  \overline{K},\ \ {\rm and/or \ \ }\ \ \ \ {\rm Decay: \ \ \ }(K, \overline{K})\stackrel{\Slash{\rm CP}}{\longrightarrow} \pi\pi(\pi)
\end{equation}  

Therefore, we are going to derive the time evolution formulae which describe the oscillation and the decays. 
The oscillation part is already done. It is the solution to the Schr\"{o}dinger equation given above. The states at time $t$, starting as $K$ and $\overline{K}$ at  $t=0$ are given: 
\begin{eqnarray}
|K(t)\rangle &=& f_+(t)|K\rangle + \frac{q}{p}f_-(t)|\overline{K}\rangle \\
|\overline{K}(t)\rangle &=& f_+(t)|\overline{K}\rangle + \frac{p}{q}f_-(t)|{K}\rangle 
\end{eqnarray}
where 
\begin{equation}
f_{\pm}=\frac{1}{2}e^{-iM_1t}e^{-\frac{1}{2}\Gamma_1t}\left[1\pm e^{-i\Delta M t}e^{\frac{1}{2}\Delta \Gamma  t}\right]
\end{equation}

Now the decay part. The decay amplitude of $K/\overline{K}$ to given final state $f$ ($f=\pi\pi$ or $\pi\pi\pi$) can be expressed by the matrix element with effective Hamiltonian with $\Delta S=1$: 
\begin{equation}
A(f)=\langle f|{\mathcal H}_{\Delta S=1}|K^0\rangle, \quad 
\overline{A}(f)=\langle f|{\mathcal H}_{\Delta S=1}|\overline{K}^0\rangle
\end{equation}
Then, the decay width of the state which was $K^0$ and $\overline{K}^0$ at $t=0$ reads: 
 \begin{eqnarray}
 \Gamma(K^0(t)\to f)&\propto& e^{-\Gamma_1t}|A(f)|^2\left[K_+(t)+K_-(t)\left|\frac{q}{p}\right|^2|\overline{\rho}(f)|^2+2\Re\left[L^*(t)\left(\frac{q}{p}\right)\overline{\rho}(f)\right]\right]
\nonumber \\ 
 \Gamma(\overline{K}^0(t)\to f)&\propto& e^{-\Gamma_1t}|\overline{A}(f)|^2\left[K_+(t)+K_-(t)\left|\frac{p}{q}\right|^2|{\rho}(f)|^2+2\Re\left[L^*(t)\left(\frac{q}{p}\right){\rho}(f)\right]\right]
 \nonumber
 \end{eqnarray}
where 
\begin{eqnarray}
\overline{\rho}(f)&\equiv&\frac{\overline{A}(f)}{A(f)}\equiv\frac{1}{\rho(f)} \\ 
|f_{\pm}(t)|^2&=&\frac{1}{4}e^{-\Gamma_1t} K_\pm (t) \\ 
f_{-}(t)f_{+}^*(t)&=&\frac{1}{4}e^{-\Gamma_1t} L^* (t) \\ 
K_{\pm}(t) &=& 1+e^{\Delta \Gamma}\pm 2e^{\frac{1}{2}\Delta \Gamma t}\cos \Delta Mt \\
L^*(t) &=& 1-e^{\Delta \Gamma}+ 2ie^{\frac{1}{2}\Delta \Gamma t}\sin \Delta Mt 
\end{eqnarray}
The CP violation manifests itself as: 
\begin{equation}
{\mathcal{A}}=\frac{  \Gamma(\overline{K}^0(t)\to f)-\Gamma(K^0(t)\to f)}{  \Gamma(\overline{K}^0(t)\to f)+\Gamma(K^0(t)\to f)}\neq 0
\end{equation}

\subsection{The three types of CP violation}
In this section, we learn the three types of CP violating processes: 
\begin{itemize}
\item Direct CP violation (no-oscillation) 
\item Flavour specific mixing CP violation 
\item Flavour non-specific mixing CP violation (time dependent CP violation)
\end{itemize}

\noindent 
\underline{\bf Direct CP violation (no-oscillation):} \hspace*{0.1cm}\\
No-oscillation means $\Delta M=0, \Delta\Gamma =0$ then, we have $K_-(t)=L(t)=0$. 
In this type, CP violation occurs only through the decay: 
\begin{equation}
|A(f)|\neq |\overline{A}(\overline{f})|
\end{equation}
The CP asymmetry is given as: 
\begin{equation}
{\mathcal{A}}=\frac{|\overline{A}(\overline{f})|^2-|A(f)|^2}{|\overline{A}(\overline{f})|^2+|A(f)|^2}
=\frac{|\overline{\rho}(\overline{f})|^2-1}{|\overline{\rho}(\overline{f})|^2+1}
\end{equation}
It should be noted that non-zero CP asymmetry ${\mathcal{A}}\neq 0$ occurs only when $|\overline{\rho}|\neq 1$ ($\arg(\overline{\rho})\neq 0$ is not sufficient!). \hspace*{0.2cm}

\noindent 
\underline{\bf Flavour specific mixing CP violation :} \hspace*{0.1cm}\\
Let's consider the semi-leptonic decay, e.g. $K^0\to X l^+\nu$ or $\overline{K}^0\to X l^-\overline{\nu}$. 
Note that at the level of quark and leptons, these decays come from $\overline{s}\to\overline{u}W^+(\to l^+\nu)$ and $s\to u W^-(\to l^-\overline{\nu})$, respectively. In such a decay mode, the initial state and the final state have one to one correspondence: tagging of the final state flavour (or lepton charge) tells whether the initial state was $K^0$ or $\overline{K}^0$. Defining the decay amplitude as: 
\begin{equation}
A_{SL}\equiv |A(Xl^+\nu)|=|\overline{A}(Xl^-\overline{\nu})| 
\end{equation}
 (note i) this equality comes from CPT invariance and ii) $|A(Xl^-\overline{\nu})|=|\overline{A}(Xl^+{\nu})| =0$), we find the decay rates for the state which was $K^0$ or $\overline{K}^0$ at $t=0$ read: 
 \begin{eqnarray}
 \Gamma(K^0(t)\to l^+X)&\propto& e^{-\Gamma_1t} K_+(t)|A_{SL}|^2 \\
 \Gamma(K^0(t)\to l^-X)&\propto& e^{-\Gamma_1t} K_-(t)\left|\frac{q}{p}\right|^2|A_{SL}|^2 \\
 \Gamma(\overline{K}^0(t)\to l^-X)&\propto& e^{-\Gamma_1t} K_+(t)|A_{SL}|^2 \\
 \Gamma(\overline{K}^0(t)\to l^+X)&\propto& e^{-\Gamma_1t} K_-(t)\left|\frac{p}{q}\right|^2|A_{SL}|^2 
 \end{eqnarray}
where the wrong sign processes (the second and the fourth lines) come from the $K^0\leftrightarrow \overline{K}^0$ oscillation. The CP asymmetry is given as: 
\begin{equation}
{\mathcal{A}}=\frac{|p/q|^2-|q/p|^2}{|p/q|^2+|q/p|^2} =\frac{|p/q|^4-1}{|p/q|^4+1}
\end{equation}
which does not depend on the time. 
\hspace*{0.2cm}

\noindent 
\underline{\bf Flavour non-specific mixing CP violation (time dependent CP violation) :} \hspace*{0.1cm}\\
For this type of CP violation to be measured, we utilize very special kinds of final state: the final state to which both $K^0$ and $\overline{K}^0$ can decay. The CP eigenstate $CP|f_{\pm}\rangle =\pm |f_{\pm}\rangle$ falls into this category.  
Indeed, the $\pi\pi$ final states are such a case: 
\begin{equation}
K^0\to \pi^+\pi^-, \overline{K}^0\to \pi^+\pi^-, \quad  
K^0\to \pi^0\pi^0, \overline{K}^0\to \pi^0\pi^0
\end{equation}
In general, both $|\overline{\rho}(f)|\neq 1$ and $q/p\neq 1$ can occur. Just for simplicity, we present the result for $|\rho(f)|= 1$ and $|q/p|= 1$, 
\begin{equation}
{\mathcal{A}}=\frac{2\sin(\arg{q/p}+\arg{\overline{\rho}})e^{\frac{1}{2}\Delta\Gamma t}\sin \Delta Mt}{1+e^{\Delta \Gamma t}+\cos(\arg{q/p}+\arg{\overline{\rho}})[1-e^{\Delta \Gamma t}]}
\end{equation}
Thus, the non-zero CP asymmetry will occur when $\arg{q/p}+\arg{\overline{\rho}}=\neq$ and $\Delta M\neq 0$. The asymmetry depends on the time in this case. 
We will come back to this type of CP asymmetry later on the $B$ meson system. 

\subsection{CP violation in $B\overline{B}$ system}
The discovery of the CP violation in $K$ system is helped by the (accidental) fact that the two (supposed-to-be) eigenstates $K_S$ and $K_L$ have very different life time, which allowed us to realize that $K_L $ (CP-odd state) decayed to $\pi\pi$ (CP even state). In the $B$ meson system,  of two $B$ states both have very short life time. Thus, we need some strategy to identify whether the initial was $B$ or $\overline{B}$. The most common way to achieve this task is the following: 
\begin{itemize}
\item $t=0$:  $B$ and $\overline{B}$ are pair-produced from $e^+e^-$ collision (in this way, the $B$$\overline{B}$ is produced in a $C$ odd configuration). 
\item $t=t_1$: one of $B$ or $\overline{B}$ decay semi-leptonically. As presented in the previous section, if the final state contained $l^{-(+)}$, then, the particle that decayed was $\overline{B}(B)$. Due to the quantum-correlation, if $l^{-(+)}$ is detected, the other particle which hasn't decayed yet should be $(B)\overline{B}$. 
\item $t=t_2$: Then, this remaining particle decays to the CP eigenstate, which is common for $B$ and $\overline{B}$. Between $t=t_1$ and $t=t_2$, this particle oscillate between $B$ and $\overline{B}$. 
\end{itemize}
 
\begin{figure}
\begin{center}
\resizebox{0.7\columnwidth}{!}{%
  \includegraphics{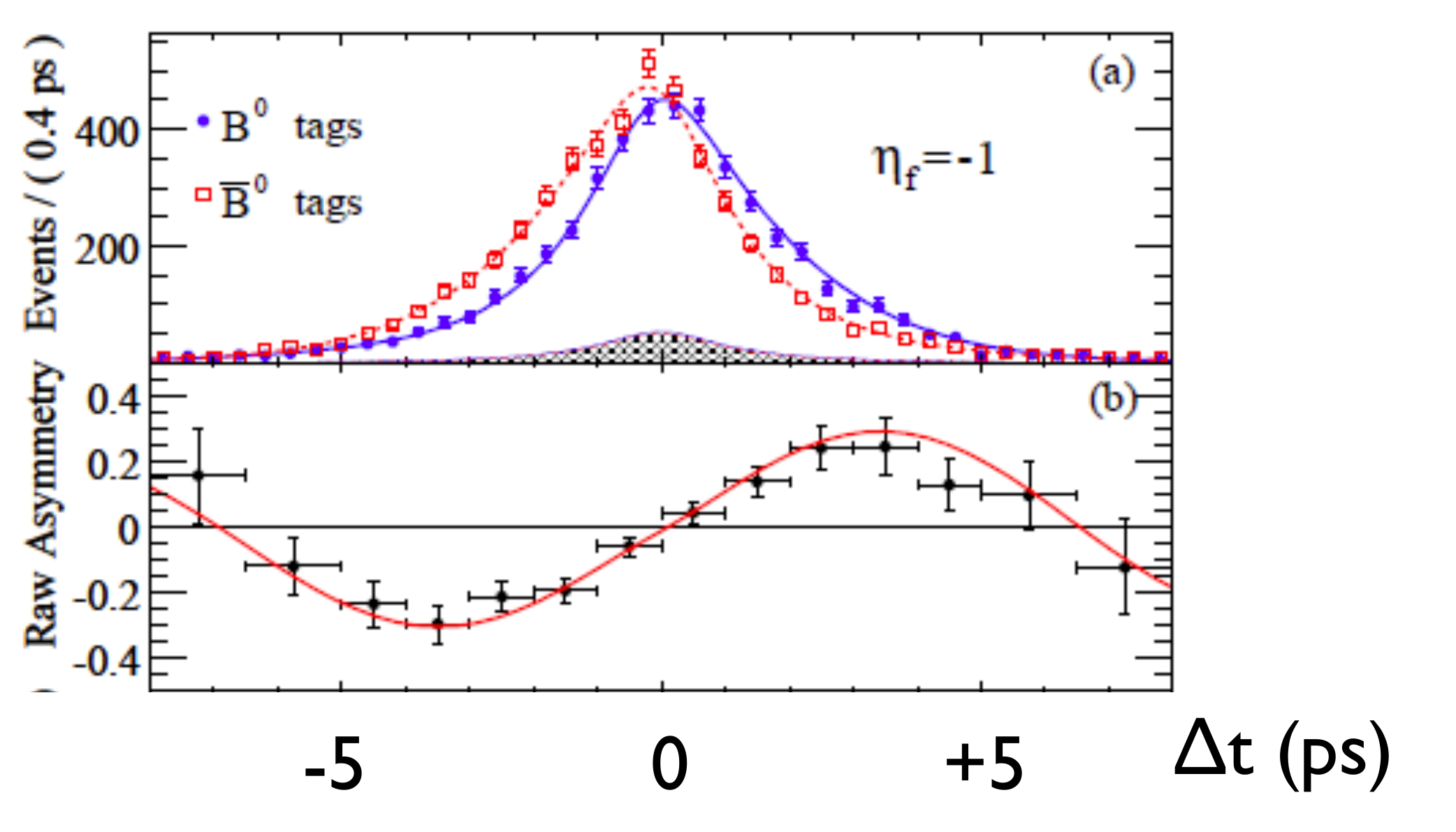} }
\end{center}
\caption{Time dependence of the $B-\overline{B}$ oscillation. }
\label{SM-fig:1}       
\end{figure}

The decay rate  at $t=t_2$ for the processes where we observe $l^{\pm}$ at $t=t_1$  
can be written as: 
\begin{eqnarray}
\Gamma(B^0(t_2)\to f)&\propto & e^{-\Gamma_B(t_2-t_1)}|A(B^0\to f)|^2[1-\Im( \frac{q}{p}\overline{\rho}(f))\sin(\Delta M_B(t_2-t_1))] \nonumber \\
\Gamma(\overline{B}^0(t_2)\to f)&\propto & e^{-\Gamma_B(t_2-t_1)}|A(B^0\to f)|^2[1+\Im (\frac{q}{p}\overline{\rho}(f))\sin(\Delta M_B(t_2-t_1))] \nonumber 
\end{eqnarray}
where $\overline{\rho}=1$ is assumed for simplicity and also $\Delta \Gamma_B=0$ is assumed, which is close to the truth from the observation. If CP is violated $q/p\neq 1$, we should observe different time dependence for these two processes. Indeed, experiment has observed a clear difference between this two and CP violation was confirmed at B factory experiments in 2001 with the final state $f=J/\psi K_S$ (see Fig.~\ref{SM-fig:1} top). It was 35 years after the first discovery of CP violation in $K$ decay. In this channel, the time dependent of the asymmetry behaves as: 
\begin{equation}
{\mathcal{A}}=\frac{ \Gamma(\overline{B}^0(t)\to f)-\Gamma(B^0(t)\to f)}{  \Gamma(\overline{B}^0(t)\to f)+\Gamma(B^0(t)\to f)}=\Im \frac{q}{p}\overline{\rho}(f)\sin \Delta M_Bt 
\end{equation}
where $t=t_2-t_1$. 
Comparing to the kaon system, the CP violation in $B$ system appeared to be large $\Im \frac{q}{p}\overline{\rho}(J/\psi K_S)\simeq 0.67$. 

\section{CP violation in SM: unitarity triangle}
Now that we have enough evidences of CP violation in nature, both in $K$ and $B$ system. In fact, by now, not only in $K\to \pi\pi(\pi)$ and $B\to J/\psi K_S$ processes, but also CP violation has been observed in many different decay channels. The CP violation for these two channels indicates namely $\arg (q/p)\neq 0$ in $K$ and $B$ system. There are also relatively large {\it Direct CP violation} observed in various channels (such as $B\to \pi\pi$), which indicates $|\overline{\rho}|\neq 1$ in those channels. A hint of observation of the {\it flavour-specific CP violation} is also reported ($|q/p|\neq 1$) in $B_s$ system but the experimental result is not precise enough yet. We will discuss on this issue later in this lecture. 

In this section, we discuss {\it where the complex phase comes from} in the SM in order to have $\arg (q/p)\neq 0$. In the model building point of view, it is not easy to  incorporate a complex parameter to the theory, namely because, the CP violation is observed only in the $K$ and $B$ systems but nowhere else. Most strong constraint for introducing complex phase to the theory comes from the non-observation of the electric-dipole moment (EDM) of leptons and neutrons. As we have discussed before, the Yukawa matrix contains free parameters of SM and can be no-Hearmitian. The observable of the Yukawa matrix, the CKM matrix, is only constrained to be unitary, thus can contain a complex number.  The CKM matrix is the coupling for the flavour changing charged current, thus, it is ideal to generate CP violation only in the flavour non-diagonal sectors. 

\subsection{Kobayashi-Maskawa ansatz}
The fact that the CKM matrix contains complex phases does not necessarily mean that they generate observable CP violation, since some phases can be absorbed by the redefinition of the field. 
In 1973, Kobayashi and Maskawa investigated this question. A general $n\times n$ unitary matrix contains $2n^2-(n+(n^2-n))=n^2$ real parameters. The phases of the quark fields can be rotated freely, $\psi \to e^{i\phi}\psi$ (applying separately for up-type and down-type quarks),  while one overall phase is irrelevant. Thus, we can rotate $2n-1$ phases by this. As a result, we are left with $n^2-(2n-1)=(n-1)^2$ real parameters. Among these parameters, we subtract the number of the rotation angles, which is the number of the real parameter in $n\times n$ orthogonal matrix $\frac{1}{2}n(n-1)$. 
As a result, the number of the independent phase in CKM matrix is: $\frac{1}{2}(n-1)(n-2)$. 
Kobayashi and Maskawa concluded that {\it in order for CP to be broken through CKM matrix, third generation of quarks is necessary}. In 1973 when they wrote this paper, there were only three quarks confirmed (up, down and strange) with a speculation of the fourth quark (charm). The prediction of further two quarks was rather bold. However, indeed, the $J/\psi$ (a charm anti-charm bound state) was discovered in 1974. The third generation lepton $\tau$ was seen in 1975 and confirmed in 1977. Also in 1977, the fifth quark, bottom was discovered. For the sixth quark, top, are needed to wait until 1994.  
Now the Kobayashi and Maskawa mechanism is a part of the SM. As we see in the following, all the observed CP violations can be explained by the single phase in the CKM matrix {\it at a certain level}. Therefore, it is believed that this phase is the dominant source of the observed CP violation. 

\subsection{The unitarity triangle}
As we have repeated, the CKM matrix is restricted  by theory only to be unitary. It contains four free parameters (three rotation angles and one phase), which should explain all observed flavour changing and non-changing phenomena including CP violating ones. Thus, the test of the unitarity of the CKM matrix is a very important task in particle physics. For this purpose, let us first inroduce the most commonly used parameterization. 

First we write the $3\times 3$ unitary matrix as product of three rotations ordered as:
\begin{eqnarray}
{\bf V}&=&\omega(\theta_{23},0)\omega(\theta_{13},-\delta)\omega(\theta_{12},0) \\
&=& \left(\begin{array}{ccc}
c_{12}c_{13} & s_{12}c_{13} & s_{13}e^{-i\delta} \\
-s_{12}c_{23}-c_{12}s_{23}s_{13}e^{i\delta} & c_{12}c_{23}-s_{12}s_{23}s_{13}e^{i\delta} & s_{23}c_{13} \\
s_{12}s_{23}-c_{12}c_{23}s_{13}e^{i\delta} & -c_{12}s_{23}-s_{12}c_{23}s_{13}e^{i\delta } & c_{23}c_{13} \end{array}\right) \phantom{+{\mathcal{O}}(\l{4})}
\end{eqnarray}
Now we re-define the parameters (Wolfenstein's parameterization): 
\begin{equation}
\sin{\theta_{12}}=\lambda, \quad 
\sin{\theta_{13}}=A \sqrt{\rho^2+\eta^2} \lambda^{3}, \quad 
\sin{\theta_{23}}=A\lambda^{2} 
\end{equation}
Then, realizing that the observed CKM matrix elements follow some hierarchy, we expand in terms of $\lambda(\simeq 0.22)$: 
\be
\def\l#1{\lambda^{#1}}
{\bf V}=\left(\begin{array}{ccc}
1-\frac{1}{2}\l{2} & \lambda & A\sqrt{\rho^2+\eta^2}e^{-i\delta}\l{3}  \\
-\lambda & 1-\frac{1}{2}\l{2}& A\l{2}\\
A(1-\sqrt{\rho^2+\eta^2}e^{i\delta})\l{3} 
& -A\l{2} & 1 
\end{array}\right) +{\mathcal{O}}(\l{4})
\ee

\begin{figure}
\begin{center}
\resizebox{0.7\columnwidth}{!}{%
  \includegraphics{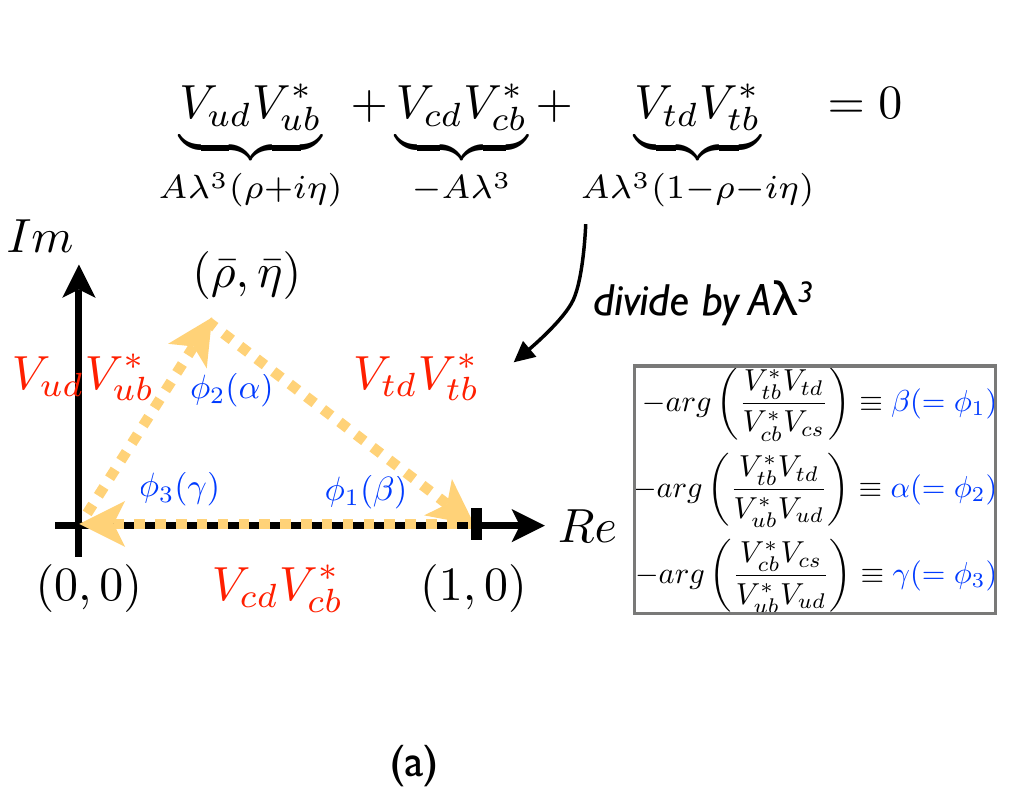}}\\
\resizebox{0.45\columnwidth}{!}{%
  \includegraphics{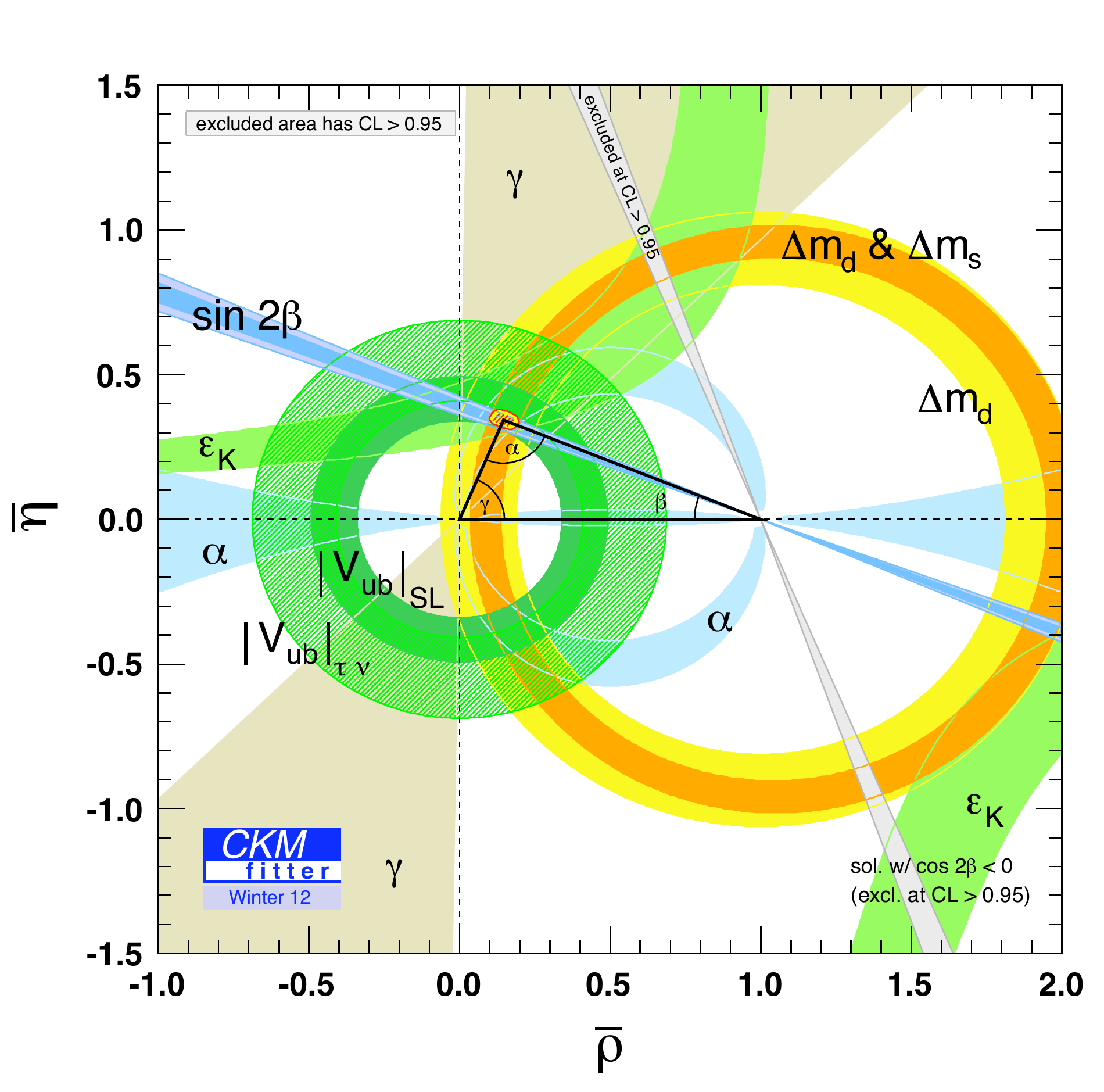} }\ \ \ 
\resizebox{0.45\columnwidth}{!}{%
  \includegraphics{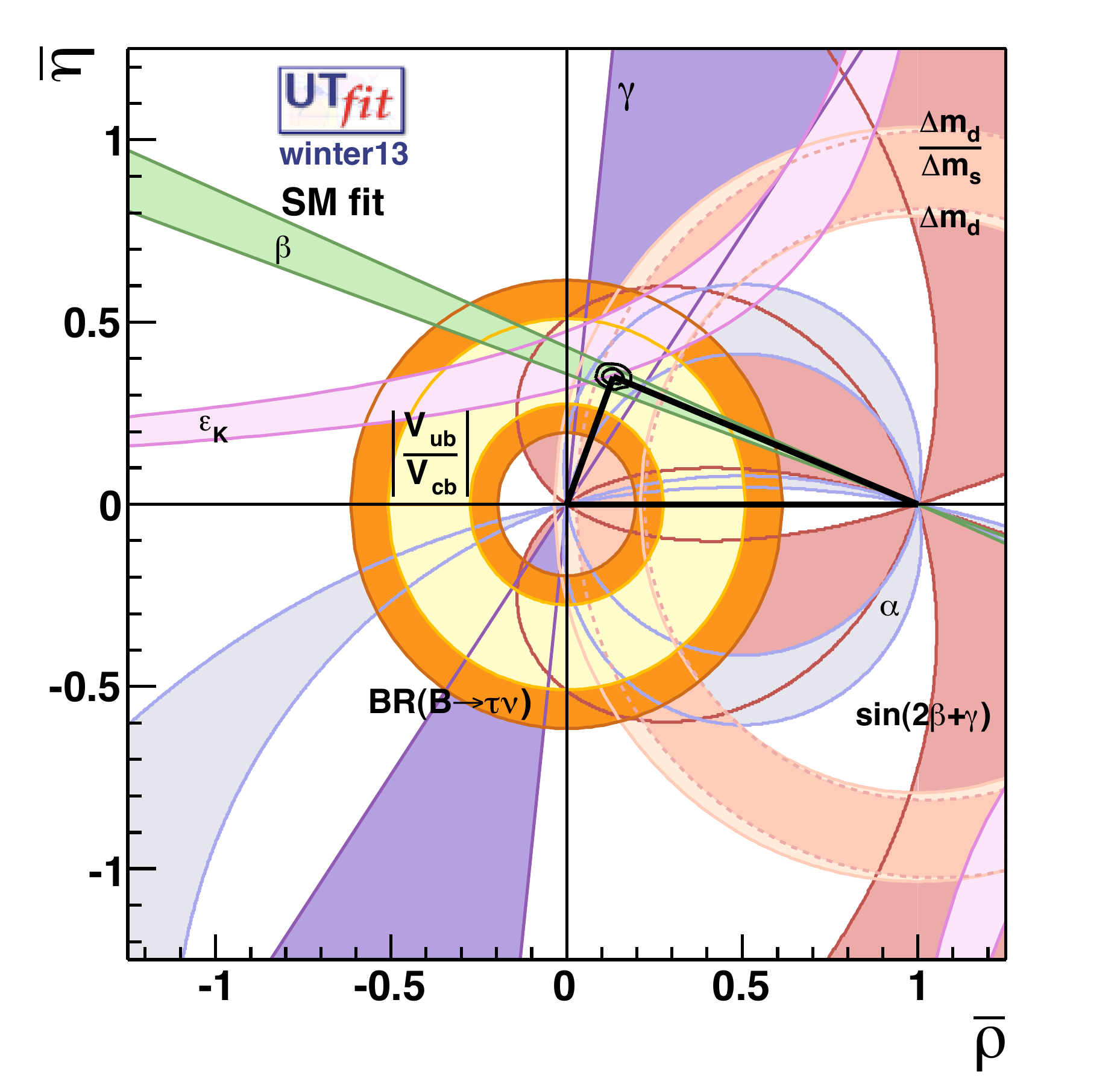} }
\caption{(a) The unitarity triangle. (b) and (c) The current situation of the unitarity triangle constraints from various flavour observables.}
\label{SM-fig:3}       
\end{center}
\end{figure}

In testing whether all the observed FCNC and CP violating phenomena can be explained by the CKM matrix (unitary matrix which can be written in terms of three rotation angles and one complex phase), the so-called unitarity triangle is often useful. The unitarity triangle represents one of nine unitarity conditions, the one that contains the matrix elements relevant to B physics: 
\begin{equation}
V_{ud}^*V_{ub}+V_{cd}^*V_{cb}+V_{td}^*V_{tb}=0
\end{equation} 
Assuming that each term is a vector in the complex plane (known as the $\bar{\rho}=\bar{\eta}$ plane), we can draw a triangle (see Fig.~\ref{SM-fig:3} a). 
We measure {\it independently} the sides and the angles of this triangle to test, in particular, whether the triangle closes or not. The latest result is shown in Fig.~\ref{SM-fig:3} b and c. Let us first look at the measurements of two sides, $|V_{ub}|$ (left side) and $\Delta M_d/\Delta M_s$ (right side). The overlap region of theses two measurements determine roughly the position of the apex of the triangle. One can see that the triangle is {\it not  flat} from these constraints. The one of the three angles, $\beta(=\phi_1)$ is measured very precisely at the B factories through the observation of the $B_d$ oscillation. And this angle is measured as $(21.7\pm 0.64)^{\circ}$, which is indeed rather large. The right side of the triangle drawn by using this value of $\beta(=\phi_1)$, the allowed bound passes through the allowed range from the $|V_{ub}|$ and $\Delta M_d/\Delta M_s$ measurements. Moreover, the overlapping region from these three measurements has also an overlap with the allowed range from the $K$ oscillation measurement, $\epsilon_K$. 

The apex of the triangle determined from various measurements is constrained to be  in {\it a small region}, indicating that these phenomena can be explained by the four free parameters of the SM which are in the CKM matrix. In particular, the success of the KM mechanism is manifested by the CP violation in the $K$ and the $B$ systems being explained by the single complex phase in the CKM matrix. 

However, the Fig.~\ref{SM-fig:3} b and c apparently show that the whole program of verifying the unitarity of the CKM matrix has not been finished yet. The remaining two angles, $\alpha(=\phi_2)$ and $\gamma(=\phi_3)$, have not been measured as precisely as $\beta(=\phi_1)$. Indeed, experimentally, the LHCb experiment has an ability to determine $\gamma(=\phi_3)$ through e.g. $B\to D^{(*)}K$ modes at a higher precision. It will be interesting to see if the right side drawn by using a more precise $\gamma(=\phi_3)$ measurement in the future will still pass through the apex regions allowed by the other measurements. We should like to draw the attention to a {\it subtle tension} appearing in the Fig.~\ref{SM-fig:3} b and c: the overlap region among $|V_{ub}|$, $\Delta M_d/\Delta M_s$ and $\beta(=\phi_1)$. For now, these three bounds have an overlapping region as discussed above. However, the latest determination of $|V_{ub}|$ from the measurement of the branching ratio of $B\to\tau\nu $ turned out to be slightly higher than the ones determined from the semi-leptonic $b\to ul\nu$ decays. If this tendency remains and the $|V_{ub}|$ value shifts towards a larger value, then, the overlap region with $\beta(=\phi_1)$ could be lost. The super B factory, which are now approved project for B physics, has an ability to measure the $B\to \tau\nu$ branching ratio at a much higher precision. Thus, it will not be too long before the hint of this tension will be revealed. Finally, we should also mentioned that the errors indicated in the Fig.~\ref{SM-fig:3} b and c contain not only the experimental ones but also the theoretical ones, namely coming from the hadronic uncertainties. And in particular for $|V_{ub}|$ and $\Delta M_d/\Delta M_s$, the theoretical uncertainties are the dominant sources of the error. Thus, in order to achieve a high prevision in determining these parameters, a reduction of the theoretical uncertainty is the most essential. Progresses in various theoretical methods based on QCD, in particular, Lattice QCD, are key for this goal. 

\section{CP violation in the $B_s$ system: search for physics beyond the SM}
\subsection{The $B_s$ oscillation}
We can derive the $B_s$ oscillation formulae in the same way as $B_d$ system. 
Experimentally, the following quantities are measured: 
\begin{eqnarray}
& \Delta M_s\equiv M_2-M_1=-2|M_{12}|, \quad
\Delta \Gamma_s \equiv \Gamma_1-\Gamma_2=2|\Gamma_{12}|\cos\zeta_s & \label{eq:5}\\
&\frac{q}{p}\simeq  e^{-i\phi_s}\left(1+\frac{\Delta\Gamma_s}{2\Delta M_s}\tan\zeta_s\right), \quad 
\left|\frac{q}{p}\right|\simeq  1+\frac{\Delta\Gamma_s}{2\Delta M_s}\tan\zeta_s \label{eq:6_app}
&
\end{eqnarray}
where the phases are defined as 
\begin{equation}
\phi_s\equiv \arg[M_{12}], \quad \zeta_s\equiv \arg[\Gamma_{12}]-\arg[M_{12}] \label{eq:7_app}
\end{equation}
$\Delta M_s$ is measured by Tevatron rather precisely, $\Delta M_s=(17.77\pm 0.19\pm 0.07)$ ps$^{-1}$. Recently, many progresses have been made for determining the CP violating phase $\phi_s$. In SM, this phase is related to $\phi_s=\beta_s\equiv -arg[V_{tb^*}V_{ts}/V_{cb^*}V_{cs}]$. The SM phase is known to be very small $\beta_s\simeq 2^{\circ}$, while the LHCb experiment has an ability to reach to this level measuring the $B_s$ oscillation with the $b\to c\bar{c}s$ decay channels such as $B_s\to J/\psi\phi$ or $B\to J/\psi f_0$. It should be noted that different from the $B_d$ system, $\Delta\Gamma_s/\Delta M_s$ is non-negligible. Thus, the precise determination of $\phi_s$ requires the information of $\Delta\Gamma_s$. Also note that in the case of the $J/\psi \phi$ final state, the angular analysis is required to decompose different polarization state which have different CP. 
We also have another observable, the $B_s$ oscillation measurement with the lepton final state, namely the di-lepton charge asymmetry $A_{sl}$, which determines $|q/p|$. 

\begin{figure}
\resizebox{1\columnwidth}{!}{%
  \includegraphics{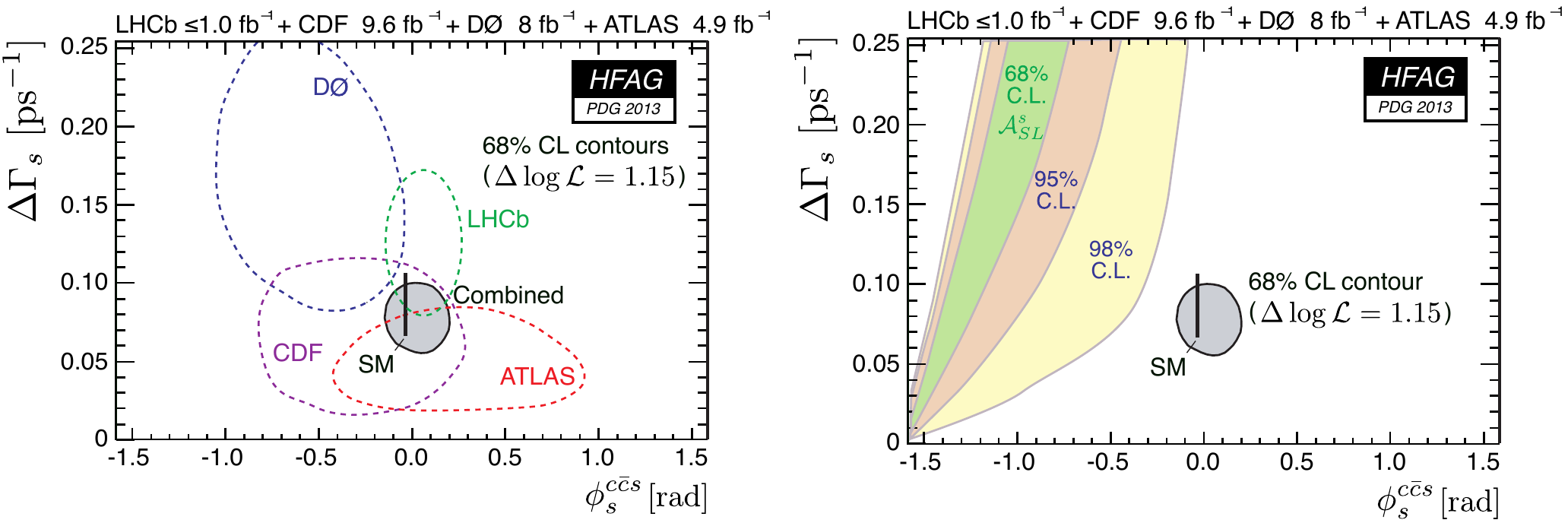} }
\caption{The current experimental bounds on the $B_s$ oscillation parameters.}
\label{fig:1_app}       
\end{figure}

The constraints on $\phi_s$ and $\Delta \Gamma_s$  from the CDF and the D0 collaboration averaged by HFAG are presented in Fig.~\ref{fig:1_app}. On the left figure, the constraint from the $B_s$ oscillation measurement using the $Bs\to J/\psi \phi$ and $Bs\to J/\psi f_0$ final states (contours) and the constraint from the di-lepton charge asymmetry measurement (curved bound) are separately plotted while on the right figure, the combined constraints from these two is presented.  It is important to notice that $\Delta \Gamma_s$ is a function of the phase $\zeta_s$ as shown in Eq.~\ref{eq:6_app} where $\zeta_s$ is related to $\phi_s$ as in Eq.~\ref{eq:7_app}. In particular, as the $\Gamma_{12}$ is the imaginary part of $B_s-\bar{B}_s$ box diagram, which comes from the up and  the charm quark contributions, it is real, unless a new physics contributed to the imaginary part with a non-zero CP violating complex phase. In the case of $\arg\Gamma_{12}=0$, we have a physical  region is such that the $\Delta \Gamma_s$ always decreases from the SM value when  $|\phi_s|$ departs from its SM value $~0^{\circ}$. The average of $B_s\to J/\psi \phi$ and $B_s\to J/\psi f_0$ for $\phi_s$ measurement is obtained as: 
\begin{equation}
\phi_s=0.04^{+0.10}_{-0.13} 
\end{equation}
which is smaller than the previous Tevatron result ($2.3 \sigma$ deviation was announced) and closer to the SM value.

\begin{figure}
\begin{center}
\resizebox{0.45\columnwidth}{!}{%
  \includegraphics{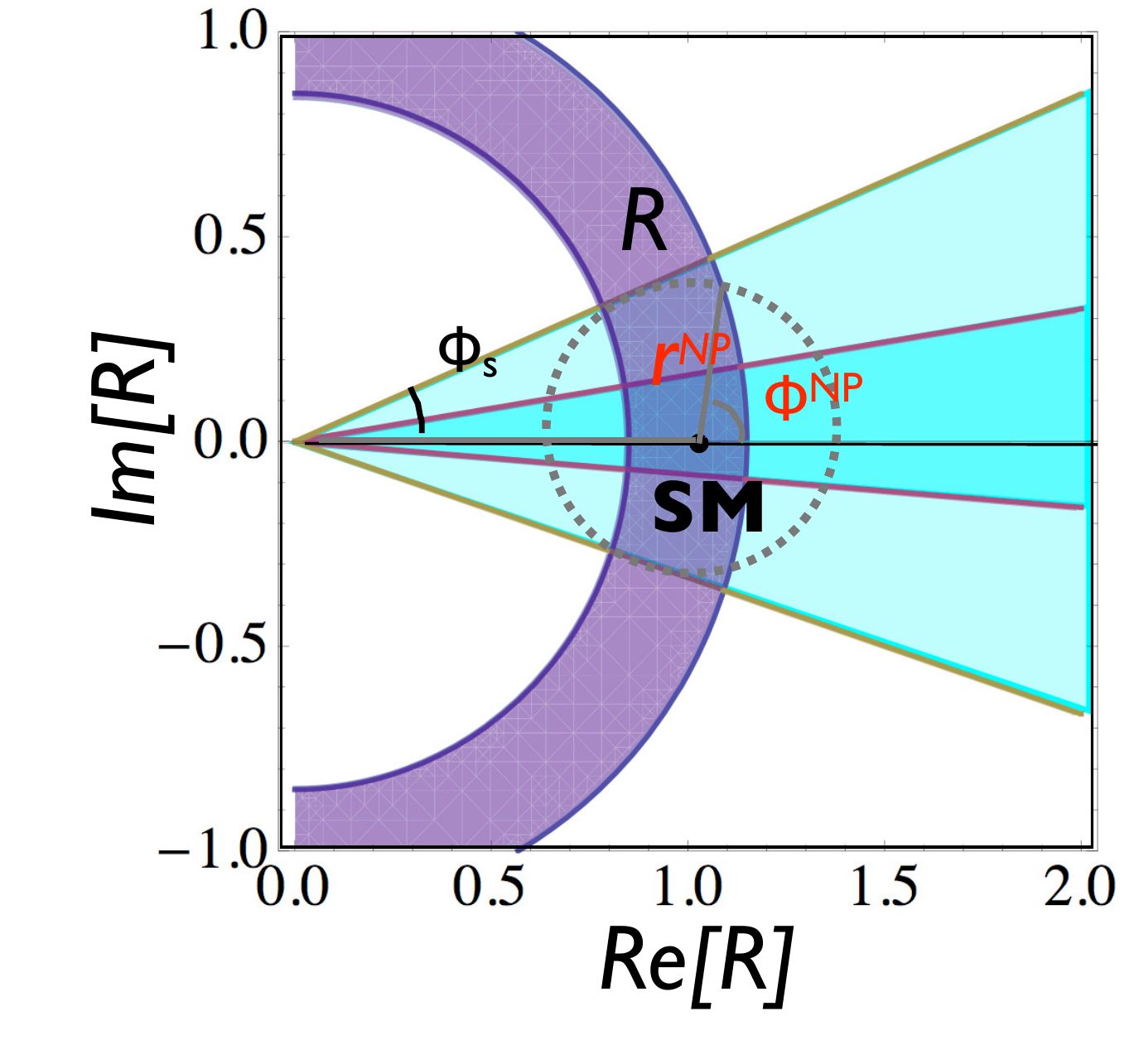} }
\end{center}
\caption{Illustration of remaining room for new physics after the $B_s$ oscillation phase measurements by LHC (see Eq.~\ref{eq:rNP}).}
\label{fig:2_app}       
\end{figure}

Next we attempt to discuss the implication of these experimental result from theoretical point of view. 
In order to clarify how large new physics effects can be still allowed after having the constraints on the $B_s$ oscillation parameters, $\Delta M_s$ and $\phi_s$, it is useful to introduce the following prameterization: 
\begin{equation}
\label{eq:rNP}
R\equiv \frac{\langle B_s^0|{\mathcal H}_{\rm eff}^{\rm SM}+{\mathcal H}_{\rm eff}^{\rm NP}|\overline{B}_s^0\rangle}{\langle B_s^0 |{\mathcal H}_{\rm eff}^{\rm SM}|\overline{B}_s^0 \rangle}\equiv1+r^{\rm NP}e^{i\phi^{\rm NP}}
\end{equation}
where NP indicates the new physics contribution. 
The Fig.~\ref{fig:2_app}  is an illustration of the allowed range of real and imaginary part of R from the $\Delta M_s$ and $\phi_s$ measurements. The purple circle represents the constraints from $\Delta M_s$ measurement. The bound includes the experimental error as well as the theoretical one, namely coming from the hadronic parameter. Here, it is illustration purpose only, thus, we assumed that the central experimental value of $\Delta M_s$ is equal to the SM value. The blue bounds represent the experimental bound coming from $\phi_s$ ($1\sigma$ and $3\sigma$ errors). The dotted-circle shows the possible new physics contribution to be added to the SM value $R\simeq 1$. What we can see this result is first, even if the experimental value for $\Delta  M_s$ is close to the SM value, the CP violating phase is allowed to be very different from the SM value ($\simeq 0$). Now that the LHCb results turned out to be close to the SM point, we can also see that 20 \% of new physics contribution can be easily accommodated even taking into account only one sigma error. As mentioned earlier, the LHCb has an ability to measure $\phi_s$ as small as the SM value ($\simeq -2^{\circ}$). Thus, there is still a plenty of hope that a new physics effect may appear in these measurements in the future.

\def\st{\scriptstyle}
\def\sst{\scriptscriptstyle}
\def\mco{\multicolumn}
\def\epp{\epsilon^{\prime}}
\def\vep{\varepsilon}
\def\ra{\rightarrow}
\def\ppg{\pi^+\pi^-\gamma}
\def\vp{{\bf p}}
\def\ko{K^0}
\def\kb{\bar{K^0}}
\def\al{\alpha}
\def\ab{\bar{\alpha}}
\def\be{\begin{equation}}
\def\ee{\end{equation}}
\def\bea{\begin{eqnarray}}
\def\eea{\end{eqnarray}}
\def\CPbar{\hbox{{\rm CP}\hskip-1.80em{/}}}
\def\lsim{\raise0.3ex\hbox{$\;<$\kern-0.75em\raise-1.1ex\hbox{$\sim\;$}}}
\def\gsim{\raise0.3ex\hbox{$\;>$\kern-0.75em\raise-1.1ex\hbox{$\sim\;$}}}
\def\stilde{\widetilde}\def\half{{1\over 2}}\def\conj{{{\rm c.c.}}}\def\sbar{\overline}
\section{Motivation to go beyond the SM}
The Standard Model (SM) is a very concise model which explains a large number observables with a very few parameters. In particular, the agreement of the electroweak precision data with  the SM predictions is quite stunning, which shows the correctness of the unified electroweak interaction with the  $SU(2)\times U(1)$ gauge symmetry, including its quantum corrections. The crucial prediction of the SM is the existence of the Higgs boson, which is at the origin of the mass of all particles in the SM. LHC has already seen some hint and the best-fitted mass around $126$ GeV is also in quite good agreement of the prediction obtained from the electroweak precision data. Furthermore, the agreement of the flavour physics is also impressive. Basically, the free parameters, three rotation angles and one complex phase in the CKM matrix can explain a large number of different experiments, including flavour changing charged/neutral currents as well as CP violating observables. 

Some ``hints of physics beyond SM" have been reported from time to time, though so far, none of them is significant enough to declare a discovery of a phenomenon beyond the SM. Then, why do we believe there is something beyond?! Indeed, the SM has a few problems. Let us list a few of them here. 
\begin{itemize}
\item {\bf Higgs naturalness problem} \\
We will see this problem more in details later on but basically this problem is related to the question of why the Higgs boson mass scale is so much lower than the Planck mass scale. The quantum corrections to the Higgs mass depend on a cut-off of the theory.  If there is no new physics scale below the Planck scale, then the quantum correction become enormous unless there is an incredible fine-tuning cancellation with the bare mass. But that would be quite unnatural. 

\item {\bf The origin of the fermion mass hierarchy} \\
In the SM there are 19 free parameters: three gauge coupling, one strong CP-violating  phase,  nine fermion masses, four parameters in the CKM matrix, and two parameters in the Higgs potential. We realize that the Yukawa interaction leads to a large number of these parameters (13 out of 19). Some people find this fact quite unsatisfactory. In particular, these values look quite random although with some  hierarchy (e.g. top quark and up or down quark have mass scale difference of order $10^5$). A symmetry in the Yukawa interaction has been searched for a while, but there has been so far no obvious solution found. 

\item {\bf The Strong CP problem} \\
Another problem concerns the one of the 19 parameters mentioned above, the strong CP-violating phase. This phase is experimentally found to be extremely small from bounds on the neutron Electric Dipole Moment (nEDM) while theoretically there is no reason why this should be so. In nature, the observed CP violation effects are all in the flavour non-diagonal sector (such as $K-\overline{K}$ or $B-\overline{B}$ oscillation) while CP violation effects in the flavour diagonal sector (such as the EDM) seems to be extremely small, if not zero. The reason for this has been searched for in relation to the conjectured flavour symmetry mentioned above. 

\item {\bf The baryon asymmetry of the universe} \\
It should also be mentioned that it is known that the CP violation is related to another problem, the baryon asymmetry of the universe, the unbalance between matter and anti-matter that occurred in the early universe.

\item {\bf Quantum theory of gravity} \\
Although it is obvious that there is a fourth interaction, the gravitational force, the SM does not incorporate this force. In fact, the quantization of gravity itself is a problem which has no solution yet. 
\end{itemize}

These problems are among the sources of the motivation to go beyond the SM. Theoretically, various types of models are proposed in order to solve one or more of the problems mentioned above. Experimentally also, tremendous efforts are payed to search for a signal beyond the SM.

\section{Flavour problems in model building beyond the SM}
One can extend the SM by introducing new fields and new interactions. The Lagrangian for these new contributions should follow certain rules (the most fundamental one, e.g. is Lorentz invariance). When adding the new terms, the most important task is to verify that these new terms do not disturb the fantastic agreement of various experimental observations with the SM predictions. 
If the new physics enters at much higher energy than the SM, then this condition could be naturally satisfied: if the currently observed phenomena are not sensitive to such high scale, the SM is valid as an effective theory. 

However, this often means that the new physics scale is extremely high (much beyond the TeV scale which can be reached by the current accelerators) or the couplings between new physics particles and the SM particles are very weak. To set a new physics scale high can be inconvenient for the new physics model building. In particular, for those models which are constructed on the motivation for Higgs naturalness problem, having another large scale much higher than the electroweak scale does not sound very preferable. Therefore, in most of the new physics models, the latter solution, to assume the flavour coupling to be very small, is applied, although it is rather artificial (comparing to the SM where such adjustment was not needed, e.g. to suppress FCNC or to explain the source of CP violation.  For example, let us consider that the $K_L-K_S$ mass difference  comes from the effective four-Fermi interaction :
\be
\frac{g^2}{M^2}\overline{\psi}_i\Gamma_\mu \psi_i \overline{\psi}_j\Gamma^\mu  \psi_j \label{eq:5f}
\ee
 If we assume the coupling to be of order 1, we find  the new physics scale to be $10^3(10^4)$ TeV (the number in parenthesis corresponds to the case when the so-called chiral-enhancement occurs) while if we assume the coupling is SM-like $g\simeq V_{td}^*V_{ts}$ then, the scale can be down to a few (few hundred)  TeV. However, to make a very strong assumption for flavour coupling is not appropriate when we are looking for a new physics signal.  

In the following, we see in some details, how the extra flavour violation and CP violation occur in the concrete models and which are the solutions. 
In general, the new physics models which encounter a serious problem from flavour physics induce tree level flavour changing neutral current (FCNC) or new sources of CP violation.

\subsection{Two Higgs Doublet Models} 
The SM consists of a single Higgs doublet which breaks the electroweak symmetry sector of the SM and gives to the particles their masses from the Yukawa interactions. On the other hand, this feature is retained also if there is more than one Higgs doublet. However, once more than one Higgs is introduced, there are extra sources of CP violation (spontaneous CP violation) and also the extra neutral Higgs can induce Flavour Changing Neutral Current (FCNC). In the following, we briefly review how these extra terms appear and the common solution to suppress them by imposing a discrete symmetry based on the so-called Natural Flavour Conservation. 

The Two Higgs Doublet Model (2HDM) is the simplest extension of the standard $SU(2)\times U(1)$ model introducing one more Higgs doublet: 
\be
\phi_1=\left(\begin{array}{c}\phi_1^+ \\ \phi_1^0\end{array}\right), \quad 
\phi_2=\left(\begin{array}{c}\phi_2^+ \\ \phi_2^0\end{array}\right)
\ee
The most general Higg potential for this model can be written as: 
\bea
V(\phi_1, \phi_2) &=& 
-\mu_1^2\phi_1^\dagger \phi_1-\mu_2^2\phi_2^\dagger \phi_2-(\mu_{12}^2\phi_1^\dagger \phi_2+h.c.)
\label{eq:V2HDM}  \\
&&+\lambda_1(\phi_1^\dagger \phi_1)^2+\lambda_2(\phi_2^\dagger \phi_2)^2
+\lambda_3(\phi_1^\dagger \phi_1\phi_2^\dagger \phi_2)+\lambda_4(\phi_1^\dagger \phi_2)(\phi_2^\dagger \phi_1)\nonumber \\
&& \frac{1}{2}\left[\lambda_5(\phi_1^\dagger \phi_2)^2+h.c.\right]
+\left[(\lambda_6\phi_1^\dagger \phi_1+\lambda_7\phi_2^\dagger \phi_2)(\phi_1^\dagger \phi_2)+h.c.\right] \nonumber
\eea
where the quadratic couplings $\mu_i$ have a mass dimension two. After imposing the Hermiticity of the potential, we find that $\mu_{12}, \lambda_{5,6,7}$ can be complex. 
After the spontaneous symmetry breaking, the Higgs fields obtain the non-zero vacuum expectation values which are invariant under $U(1)$ gauge symmetry: 
\be
\langle \phi_1\rangle =\left(\begin{array}{c}0 \\ v_1 \end{array}\right), \quad 
\langle \phi_2\rangle=\left(\begin{array}{c}0\\ v_2 e^{i\alpha}\end{array}\right)
\ee
The two VEV's, $v_{1,2}$  can have each associated phases $\delta_{1,2}$ while since the potential in Eq.~\ref{eq:V2HDM} depends only on one combination, we can rotate the basis giving one single phase $\alpha\equiv \delta_2-\delta_1$. 
Non-zero $\alpha$ induces an extra source of CP violation on top of the complex phase in the CKM matrix. Being $v_{1,2}$  the values where the potential has an stable minimum, the expectation value of the potential 
\bea
V_0&=&\mu_1^2 v_1^2+\mu_2^2 v_2^2+2\mu_{12}^2v_1v_2\cos (\delta_3+\alpha) \\
&& +\lambda_1 v_1^4 +\lambda_2 v_2^4+ (\lambda_3+\lambda_4) v_1^2v_2^2 
+ 2|\lambda_5| v_1^2v_2^2\cos (\delta_5+2\alpha ) \nonumber \\
 && +2|\lambda_6| v_1^3v_2 \cos (\delta_6+\alpha )+ 2|\lambda_7| v_1v_2^3 \cos (\delta_7+\alpha ) \nonumber
\eea
should be stable with respect to a variation of $\alpha$, i.e. $\partial V/\partial \alpha=0$. 
Note that $\delta_i$ are the complex phases of $\lambda_i$. 
This relation can be used to analyze the condition to have non-zero $\alpha$. For example, in the case when all the couplings are real, i.e. $\delta_i=0$, this relation leads to 
\be
\cos \alpha=-\frac{\lambda_6v_1^2+\lambda_7v_2^2}{4\lambda_5v_1v_2} \label{eq:min2HDM}
\ee
Thus,  CP can be broken spontaneously without an explicit CP violating phase in the Higgs coupling. 

Now, we see the Yukawa coupling of the two Higgs doublet model: 
\be
{\mathcal{L}}_Y=\sum_{ij}
\overline{\left(\begin{array}{c}U_i\\ D_i\end{array}\right)}_L
(F_{ij}\tilde{\phi_1}+ F^{\prime}_{ij}\tilde{\phi_2}) u_{jR} 
+\overline{\left(\begin{array}{c}U_i\\ D_i\end{array}\right)}_L
(G_{ij}{\phi_2}+ G^{\prime}_{ij}{\phi_1}) d_{jR} +h.c. \label{eq:Y2HDM}
\ee
where $\tilde{\phi}_i\equiv i\tau_2\phi_i^{\dagger T}=
\left(\begin{array}{c}\phi_i^{0\dagger}\\ -\phi_i^-\end{array}\right)
$ with $\tau_2$ being the Pauli matrix. After the neutral Higgs acquiring vevs, we find 
\be
{\mathcal{L}}_Y=\sum_{ij}
\overline{\left(\begin{array}{c}U_i\\ D_i\end{array}\right)}_L
m_{ij}^u u_{jR} 
+\overline{\left(\begin{array}{c}U_i\\ D_i\end{array}\right)}_L 
m_{ij}^d d_{jR} +h.c.
\ee
where 
\be
m_{ij}^u\equiv F_{ij}\langle \tilde{\phi_1}\rangle + F^{\prime}_{ij}\langle\tilde{\phi_2}\rangle , \quad 
m_{ij}^d\equiv G_{ij}{\langle\phi_2\rangle}+ G^{\prime}_{ij}{\langle\phi_1\rangle}
\ee
Now as have  done in the SM, we diagonalize the matrix $m_{ij}^{u,d}$ by the inserting unitary matrices $K_{L,R}^{U,D}$ (we need to use different matrices from left and right multiplication unless  $m_{ij}^{u,d}$ are hermitian). But here, we see a difference; when we look at the mass basis of Eq.~\ref{eq:Y2HDM} by inserting these matrices $K$, $F_{ij}^{(\prime)}$ and $G_{ij}^{(\prime)}$ are not necessarily diagonalized simultaneously, and this leads to flavour changing neutral Higgs exchanges. This can induce large FCNC contributions, which could contradict the experimental observations. 

One of the most common ways to avoid this problem is to impose the following discrete symmetry: 
\be
\phi_1 \to -\phi_1, \quad \phi_2 \to \phi_2, \quad 
d_R \to -d_R, \quad u_R \to u_R \label{eq:NFC}
\ee
which prevents $\phi_1$ from coupling to $u_R$ and $\phi_2$ to $d_R$. As a result, the FCNC can be avoided. This model is often called Type II two Higgs doublet model. As the Minimal Supersymmetric Model (MSSM) or the Peccei-Quinn models give the same phenomenological consequences, this models is the most worked type among the two Higgs doublet models. We should also note that imposing the discrete symmetry Eq.~\ref{eq:NFC} to the Higgs potential, the terms proportional to $\lambda_{6,7}$ and $\mu_{12}$ are forbidden. Then, from Eq.~\ref{eq:min2HDM}, we find $\alpha=\pm \pi/2$. This solution is equivalent to change the definition $\phi_2\to i\phi_2$, thus $\phi_1$ and $\phi_2 $ have opposite CP. Nevertheless, it is found that this solution can not lead to an observable CP violation. 

\begin{figure}
\begin{center}
 \includegraphics[width=6.5cm]%
	{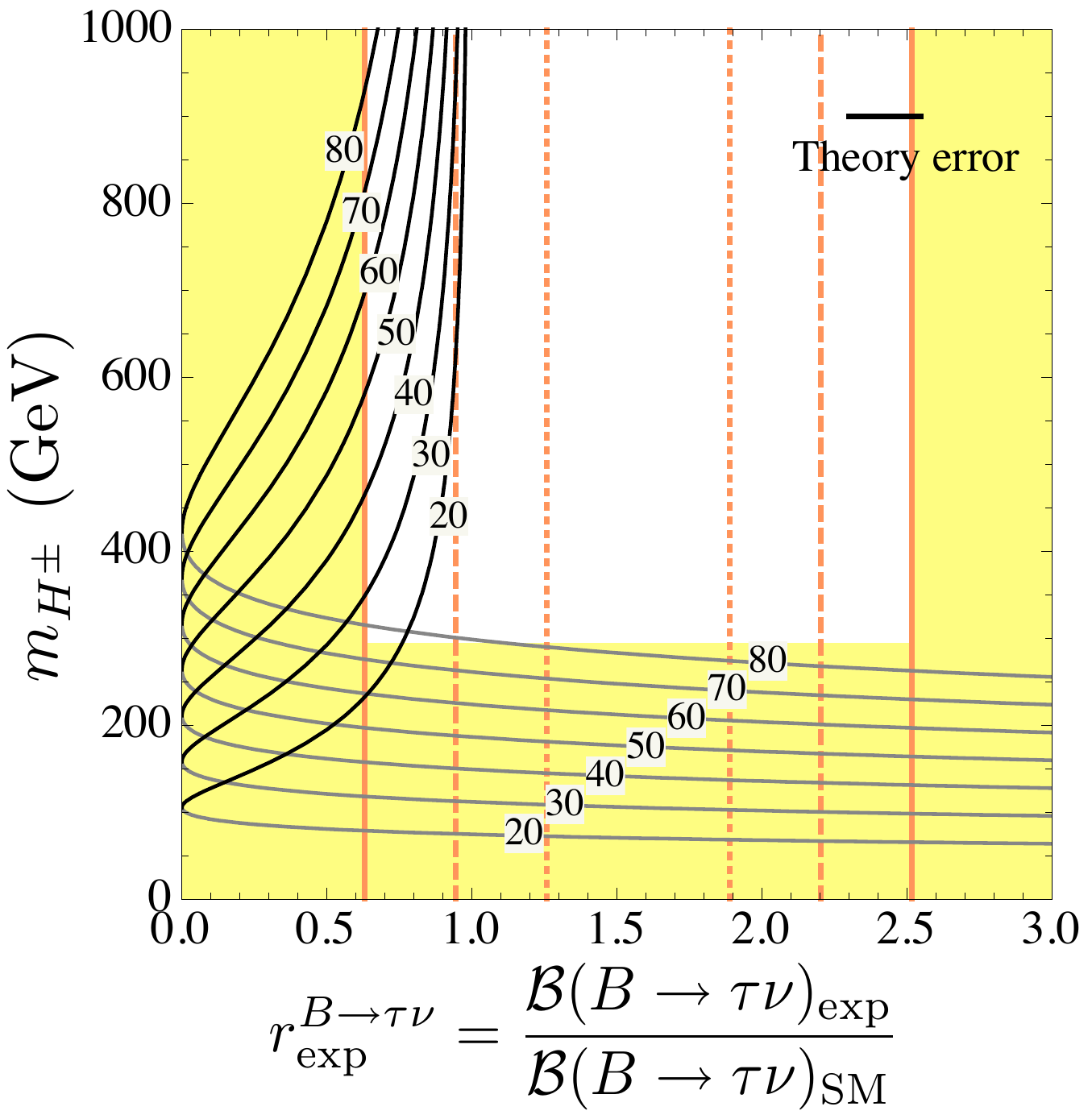}
\end{center}
 \caption{
The charged Higgs contribution to $Br(B\to \tau\nu)$ compared to the present experimental value (normalized to 
the SM prediction, $x$-axis) as a function of charged Higgs mass ($y$-axis). The vertical regions are 
excluded by the current world average of experimental value, $Br(B\to\tau\nu)_{\rm exp}=(1.15\pm0.23)\times 10^{-6}$, 
at 95\% C.L., while the $1\sigma$, $2\sigma$, $3\sigma$ errors on the same experimental value are denoted by the 
grey dotted, dashed, and solid lines, respectively. The horizontal region is excluded by the $Br(B\to X_s\gamma)$ 
measurement.     The grey and the black lines correspond to the two possible  solutions, with labels denoting the value of $\tan\beta$. The second solution 
can lead to a stronger constraint than the one from $B\to X_s\gamma$  especially for large values of $\tan\beta$.
}.
    \label{fig:2HDM}
\end{figure}

Having  suppressed the phenomenologically unacceptable  CP violation and FCNCs by the discrete symmetry,  the main effects in flavour physics are due to the charged Higgs contributions. Even though the LHC searches for the charged Higgs directly, the constraints on the property of this new particle, its mass and its couplings, come mainly indirectly from B decays. The branching ratio measurement of $B\to X_s\gamma$ constrains the mass of charged Higgs to be higher than 295 GeV. Further constraint is expected to be obtained from the branching ratio measurement of $B\to \tau\nu$ (see Fig.~\ref{fig:2HDM}). 

\subsection{The (extended) technicolor model} 
The technicolor model is one of the earliest examples of the dynamical breaking of the electroweak symmetry. The model was constructed in a close relation to QCD. In QCD, the $SU(2)$ chiral symmetry is broken spontaneously at the scale $f_{\pi}\simeq 93$ MeV, which reflects the fact that at the scale  $\Lambda_{\rm QCD}$  the QCD interaction becomes strong. Suppose, then, that there are fermions belonging to a complex representation of a new gauge group, technicolor, $SU(N_{\rm TC})$, whose coupling $\alpha_{\rm TC}$ becomes strong at $\Lambda_{\rm TC}$ around electroweak scale. Then, the relation: $M_W=M_Z\cos\theta_W=\frac{1}{2}gF_\pi$ holds, where $F_{\pi} \simeq \Lambda_{\rm TC}$ just like in QCD.  This model can nicely solve the naturalness problem since the all the produced technihadrons have masses around $\Lambda_{\rm TC}$ and they do not receive a large mass renormalization. 
However, the model is not complete unless it can provide masses to the SM fermions. For this purpose, an extension of the gauge group was proposed (Extended Technicolor Model (ETM)) that embeds flavour and technicolor into a larger gauge group. The flavour gauge symmetries are broken at a higher scale than the technicolor breaking, $\Lambda_{\rm ETC}\simeq M_{\rm ETC}/g_{\rm ETC}$ where $M_{\rm ETC}$ is the typical flavour gauge boson mass. Then, the generic fermion masses are now given by: 
\be
m_q(M_{\rm ETC})\simeq m_l(M_{\rm ETC})\simeq 2\frac{g^2_{\rm ETC}}{M^2_{\rm ETC}} \langle\overline{T}_LT_R\rangle_{\rm ETC}  \label{eq:massETC}
\ee
where $T$ is the technifermion and $\langle\overline{T}_LT_R\rangle_{\rm ETC}$ is the vacuum expectation value. However, an acquisition of the SM fermion mass by coupling to the technifermion can induce a serious flavour problem: the transition $q\to T\to q^{\prime}$ or $q\to T\to T^{\prime}\to q^{\prime}$ produce FCNC. Then, for example, the K mass difference limit and 
the $\epsilon_K$ measurement leads to the mass limit: 
\be
\frac{M_{\rm ETC}}{Re(\delta_{ds})g_{\rm ETC}}\lsim 10^3\ {\rm TeV}, \quad 
\frac{M_{\rm ETC}}{Im(\delta_{ds})g_{\rm ETC}}\lsim 10^4\ {\rm TeV},   
\ee
respectively. This value together with~\ref{eq:massETC}, we find $\Lambda_{\rm TC}$ to be 10-1000 times smaller than the electroweak scale (depending on the flavour coupling $\delta_{ds}$). 
Several solutions to this problem have been proposed. For example, the so-called "walking technicolor model" induces a large anomalous dimension which enhances the value of fermion masses by keeping  the ETC scales relatively low. This can help to reduce the FCNC for the first two generations while the top quark remains problematic, for instance in FCNC processes involving top-quark loops. A possible solution is to generate the small fermion masses by ETC, whereas the top-quark mass is dynamically generated via top condensation (Top-color assisted Technicolor model).

\subsection{Supersymmetry } 
The supersymmetric (SUSY) extensions of the SM  are one of the most popular NP models. SUSY relates fermions and bosons. 
For example, the gauge bosons have their fermion superpartners and fermions have their scalar superpartners. SUSY at the TeV scale is motivated by the fact that it solves the SM hierarchy problem. 
The quantum corrections to the Higgs mass are quadratically divergent and would drive the Higgs mass to Planck scale $\sim 10^{19}$ GeV, unless the contributions are cancelled. In SUSY models they are cancelled by the virtual corrections from the superpartners.
The minimal SUSY extension of the SM is when all the SM fields obtain superpartners but there are no other additional fields. This is the 
Minimal Supersymmetric Standard Model (MSSM). 
SUSY cannot be an exact symmetry since in that case superpartners would have the same masses as the SM particles, in clear conflict with observations. 
If supersymmetry is the symmetry of nature, the masses of the SUSY particles should be the same as their partners'. However, no candidate for SUSY particle has been detected by experiments so far.   This indicates that a more realistic model should contain  SUSY breaking terms.
Different mechanisms of SUSY breaking have very different consequences  for flavor observables. 
In complete generality the MSSM has more than one hundred parameters, most of them coming from the soft SUSY breaking terms -- the masses and mixings of the superpartners. If superpartners are at the TeV scale the most general form with ${\mathcal O}(1)$ flavor breaking coefficients is excluded due to flavor constraints. This has been called the SUSY flavor problem (or in general the NP flavor problem).

A popular solution to the SUSY flavor problem is to assume that the  SUSY breaking mechanism and the induced interactions are flavour "universal". The flavour universality is often imposed at a very high scale corresponding to the SUSY breaking mechanism. It could be at, for instance, the Planck scale
 ($\sim 10^{19}$~GeV), the GUT scale ($\sim 10^{16}$~GeV) or some intermediate scale such as the gauge mediation scale ($\sim 10^{6}$~GeV). 
The flavor breaking can then be transferred only from the SM Yukawa couplings to the other interactions through renormalization group running 
from the higher scale to the weak scale. 
As a result, the flavor breaking comes entirely from the SM Yukawa couplings (thus, an example of a concrete Minimal Flavour Violation (MFV) NP scenario). 
 Since the soft SUSY breaking terms are flavor-blind, the
 squark masses are degenerate at the high energy scale.
 The squark mass splitting occurs only due to quark Yukawa couplings, where only top Yukawa and potentially bottom Yukawa couplings are large. Thus the first two generation squarks remain degenerate to very good approximation, while the third generation squarks are split.

\begin{figure}
\begin{center}
 \includegraphics[width=10.5cm]%
	{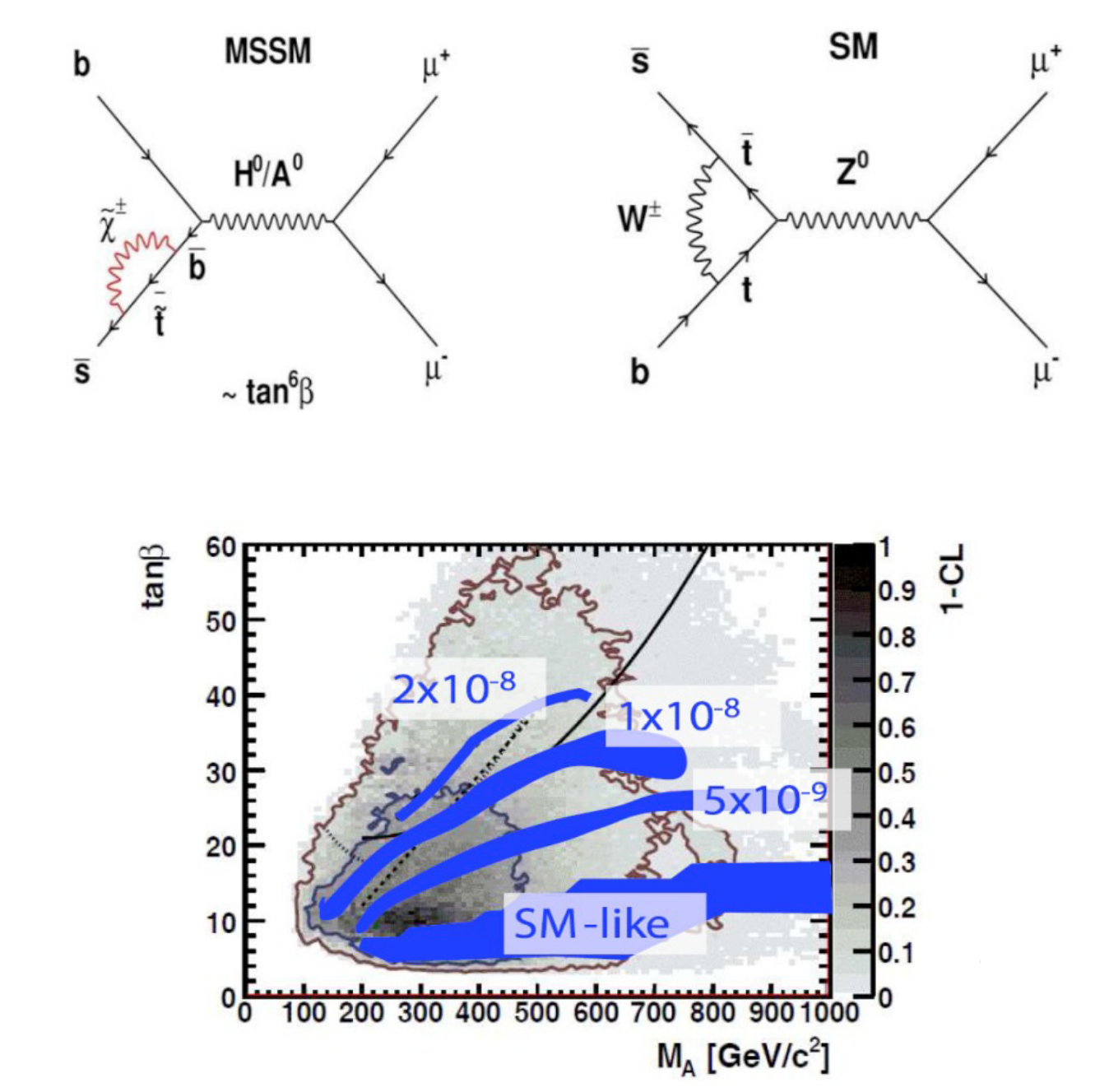}
\end{center}
 \caption{
The Feynman diagram for the $B_{s,d}\to \mu^+\mu^-$ in SM (top right) and in SUSY (top left). The constraint on the SUSY parameter obtained by using the latest LHCb result is also shown (bottom). 
}.
    \label{fig:1_Bsmm}
\end{figure}

The MFV can be extended by taking into account the large $\tan\beta$ effect. That large $\tan\beta$ scenario leads to a large new physics effects to some B physics observable and can be well tested experimentally. Most recently, the LHCb experiment has made a great progress in this scenario: observation of the $B_s\to \mu^+\mu^-$ process ($3.5 \sigma$ significance).  The $B_{s,d}\to \mu^+\mu^-$ comes from the diagram e.g. like in Fig.~\ref{fig:1_Bsmm} (top right). It is extremely rare process with branching ratios: 
\bea
Br(B_s\to \mu^+\mu^-)&=& (3.54\pm 0.30) \times 10^{-9} \\
Br(B_d\to \mu^+\mu^-)&=& (0.107\pm 0.01) \times 10^{-9} 
\eea
On the other hand, in the presence of SUSY, there is another contribution like in Fig.~\ref{fig:1_Bsmm} (top left), which can be largely enhanced by a large $\tan\beta$ factor: 
\be
Br(B_{s,d}\to \mu^+\mu^-)_{\rm SUSY}\propto  \frac{m_b^2m_{\mu}^2\tan^6\beta}{m_{A_0}^4}
\ee
Fig.~\ref{fig:1_Bsmm} (bottom) illustrate the constraint obtained by using the latest LHCb result: 
\be
Br(B_s\to \mu^+\mu^-)= (3.2^{+1.5}_{-1.2}) \times 10^{-9} 
\ee
Although the constraint depend on different models, this result excluded most of the scenario with  $\tan \beta\gsim 50$.

\section{Strategies for New Physics searches in flavour physics}
The developments of the particle physics today bore the lack of phenomena which cannot be explained in the SM. In flavour physics, many small deviations from SM (say, at the $2-3 \sigma$ level) have been reported in the past. However, definitive conclusions for those observation cannot be given so far. Therefore, the strategy in flavour physics is clear: to search for a significant enough deviation from the SM. We tackle this task from two directions, first, to improve the precision of the theoretical prediction, second, to improve the experimental precision. We should emphasize that  the latter efforts include not only experimental development, but also to propose theoretically new observables which are sensitive to new physics contributions.


Let us see the example of the CKM unitary triangle Fig.~\ref{SM-fig:3}. The measurement of the angle, e.g. $\beta$, has been improved dramatically the past 10 years since the B factories  started. However, the sides measurements ($V_{ub}, V_{cb}, V_{tb}$ etc) have not improved as much since it depends on  theoretical inputs and assumptions, namely from the strong interaction. 
In the future, there are some experimental propositions to improve the experimental measurements.  The angle measurements, such as $\beta, \gamma$, will be improved further ($\gamma$ could be determined as precise as $\beta$),  can directly be used to improve our knowledge of new physics. The usefulness of the sides measurements relies strongly on progresses in theory, in particular, the effective theory of QCD, lattice QCD or more phenomenological models.  

In Tables~\ref{table:strat1} to~\ref{table:strat4} we list the new physics sensitive observable. 
It is again amazing that all these experimental measurements agree within the theoretical and experimental errors so far. On the other hand, the LHCb experiment as well as Belle II experiment will provide us a large samples of new data in coming years. 

\begin{table}
\caption{Examples of new physics sensitive observables in B physics.  The experimental values are extracted from HFAG (mainly preparation for PDG 2013). The prospects are extracted from~\cite{BELLEIITDR} for Belle II and~\cite{LHCBTDR} for LHCb. The number for Belle II corresponds to the sensitivity at 50 ab$^{-1}$ which can be reached by the year 2023 if the commissioning starts in the year 2015 (physics run in 2016) as scheduled. For LHCb the number corresponds to the sensitivity reach at the year 2018 and the number in parethethesis is for LHCb up-grade.  }
\label{bigtable}
\begin{center}
\begin{tabular}{|c|c|c|c|}
 \hline 
 Observable & Experimental value & Prospect & Comments \\ \hline
  \begin{minipage}{3.3cm}$ S_{B\to J/\psi K_S}= \\ \sin2\phi_1(2\beta)$ \end{minipage} & $0.665\pm 0.024$&  \begin{minipage}{3.3cm} $\pm 0.012$ (Belle II) \\
$\pm 0.02 (0.007) $ (LHCb)\end{minipage}
  & 
 \begin{minipage}{4.5cm} \small The current measurement agrees with the SM prediction obtained from the $\phi_1(\beta)$ value extracted using the unitarity relation. A higher precision measurement on $\phi_1(\beta)$ together with the measurements for  the other variables in unitarity relation could reveal a new physics contribution. 
 New physics example: $bd$ box diagram and/or tree FCNC\end{minipage} \\ \hline
 $ S_{B\to \phi K_S}$ &$ 0.74^{+0.11}_{-0.13}$& \begin{minipage}{3.3cm} $\pm 0.029$ (Belle II) \\
 $\pm 0.05 (0.02)$ (LHCb)\end{minipage}& 
 \begin{minipage}{4.5cm}
\small In the year 2002, a 2-3 $\sigma$ deviation from the $S_{B\to J/\psi K_S}$ was announced though it is diminished by now. The deviation from $S_{B\to J/\psi K_S}$ is an indication of the CP violation in the decay process of $B\to \phi K_S$, which comes almost purely from the penguin type diagram. 
 New physics example: $b\to s$ penguin loop diagram \end{minipage} \\ \hline
$ S_{B\to \eta^{\prime} K_S}$ 
&$ 0.59\pm 0.07$& $\pm 0.020$ (Belle II)& 
 \begin{minipage}{4.5cm}
\small In the year 2002, a 2-3 $\sigma$ deviation from the $S_{B\to J/\psi K_S}$ was announced though it is diminished by now. The deviation from $S_{B\to J/\psi K_S}$ is an indication of the CP violation in the decay process of $B\to \eta^\prime K_S$. This decay also comes mainly from the penguin type diagram though this can be only proved by knowing the property of $\eta^{\prime}$ (quark content etc). It should also noticed that the branching ratio of this process turned out to be a few times larger than the similar charmless hadronic B decays and this could also been regarded as a hint of new physics.   
 New physics example: $b\to s$ penguin loop diagram. In particular, contributions that can induce $b\to s gg$ (followed by anomaly diagram $gg\to \eta^{\prime}$) are interesting candidates.  \end{minipage} \\ \hline
 \end{tabular}
\end{center}
\label{table:strat1}
\end{table}%

\begin{table}
\caption{Examples of new physics sensitive observables in B physics II (see caption of Table~\ref{bigtable}).}
\begin{center}
\begin{tabular}{|c|c|c|c|}
\hline
 Observable & Experimental value & Prospect & Comments\\ \hline
 $ S_{B_s\to J/\psi \phi  }=\sin2\phi_s$ & $\phi_s=0.04^{+0.10}_{-0.13}$& $\pm 0.025(0.008)$ (LHCb)& 
  \begin{minipage}{4cm}
\small The phase of $B_s$ mixing is at the order of $\lambda^4$ and very small in SM, $\sim -0.02$.  Before the LHCb started, the Tevatron data was showing a $2-3 \sigma$ deviation from the SM, which is not diminished. Since the width difference is not negligible in the $B_s$ system (unlike the $B_d$ system),  the width difference measurement has to be done simultaneously (width is less sensitive to the new physics though it is not impossible). The LHCb has an ability to reach to the precision as small as this SM value thus, there is still enough room for new physics. New physics example: CP violation in the $b\overline{s}\overline{b}s$ box diagram and/or tree FCNC. \end{minipage} \\ \hline
 $ S_{B_s\to \phi \phi  }$ & --- &$\pm 0.17 (0.03)$ (LHCb) & 
  \begin{minipage}{4cm}
\small The deviation from $S_{B_s\to J/\psi \phi(f_0)}$ is an indication of the CP violation in the decay process of $B_s\to \phi \phi$, which comes almost purely from the penguin type diagram. The analysis requires a CP state decomposition by studying the angular dependence of the  decay. The each component can include different  new physics contributions and it is complementary to the $S_{B\to \phi K_S}$ or $S_{B\to \eta^{\prime} K_S}$  measurements. In addition, the angular analysis also allows us to test the T-odd asymmetry.  New physics example: $b\to s$ penguin loop diagram
\end{minipage} \\ \hline
 \end{tabular}
\end{center}
\label{table:strat2}
\end{table}%

\begin{table}
\caption{Examples of new physics sensitive observables in B physics III (see caption of Table~\ref{bigtable}).}
\begin{center}
\begin{tabular}{|c|c|c|c|}
\hline  Observable & Experimental value & Prospect & Comments \\ \hline
 $ S_{B\to K_S\pi^0\gamma}$ & $-0.15\pm 0.20$& $\pm 0.02$ (Belle II)& 
 \begin{minipage}{4cm}
\small This is one of the golden channel for Belle II experiment where the experimental error is expected to be reduced significantly. Non-zero CP violation is the sign of the contamination of  the photon polarization which is opposite to the one predicted in SM. The theoretical error is found to be small (less than a few \%) though some authors warm a possible large uncertainties to this value.  New physics example: right handed current in $b\to s\gamma$ penguin loop diagram  \end{minipage} \\ \hline
  $ S_{B_s\to \phi \gamma }$ & --- & $\pm 0.09(0.02)$ (LHCb)& 
 \begin{minipage}{4cm}
 \small Non-zero CP violation is the sign of the contamination of  the photon polarization which is opposite to the one predicted in SM. The LHCb with its high luminosity could allow us to study this observable and it is complementary to $ S_{B\to K_S\pi^0\gamma}$ above.  New physics example: right handed current in $b\to s\gamma$ penguin loop diagram
\end{minipage}  \\ \hline
%
 $B\to K^*l^+l^-$ (low $q^2$)  & --- & \begin{minipage}{3cm} $~\sim$ 0.2 (LHCb) in \\
 $A_T^2, A_T^{\rm Im}$ 
\end{minipage}&   \begin{minipage}{4cm} 
 \small The angular distribution carry various information of new physics ($C_{7,9,10}$ and $C^{\prime}_{7,9,10}$). In particular, the low $q^2$ region is sensitive to the photon polarization of $b\to s\gamma$. \end{minipage}\\ \hline
 $B\to K_1\gamma\to (K\pi\pi)\gamma$  & --- & \begin{minipage}{3cm}
 $~\sim$ 6 \% (LHCb)\\
 $~\sim$ 18 \% (Belle II) \\
in polarization parameter $\lambda_\gamma$\end{minipage}
 &   \begin{minipage}{4cm} 
\small We can obtain the information of the photon polarization of $b\to s\gamma$ through the angular distribution of $K_1$ decay. Detailed resonance study can improve the sensitivity to the photon polarization. 
 \end{minipage}  \\ \hline
 \end{tabular}
\end{center}
\label{table:strat3}
\end{table}%

\begin{table}
\caption{Examples of new physics sensitive observables in B physics IV (see caption of Table~\ref{bigtable}). }
\begin{center}

\small

\begin{tabular}{|c|c|c|c|}

 \hline
 Observable & Experimental value & Prospect & Comments \\ \hline
 $\Delta M_{d,s}$ & \begin{minipage}{3cm}$(0.510\pm 0.004)_{B_d} $ps$^{-1}$\\$(17.69\pm 0.08)_{B_s}$ ps$^{-1}$\end{minipage} & ---&\begin{minipage}{4.6cm} \small The result is consistent to the SM prediction though it depends strongly on the $|V_{td, ts}|$. The error is dominated by the theory mainly coming from the $f_{B_{d,s}} $ and $B$ parameters. New physics example: $b\overline{d}\overline{b}d, b\overline{s}\overline{b}s$ box diagram and/or tree FCNC \end{minipage} \\ \hline
 $Br(B\to \tau \nu)$ & $(1.15\pm 0.23)\times 10^{-6}$ & $\pm$ 6\% (Belle II) & \begin{minipage}{4.6cm}
\small Up to the year 2010, the world average value was 2-3 $\sigma$ higher than the SM prediction. The SM value depends on $|V_{ub}|$ and $f_B$. New physics example: charged Higgs. 
 \end{minipage} \\ \hline
 ${\mathcal{R}}=\frac{B\to D^{(*)}\tau \nu}{B\to D^{(*)}l \nu}$ &
 \begin{minipage}{3cm}
$(0.440\pm 0.057\pm 0.042)_D $\\
$(0.332\pm 0.024\pm 0.018)_{D^*} $
 \end{minipage}& 
 \begin{minipage}{3cm}
 $\pm$ 2.5\% for Br of $D^0\tau\nu$ \\
  $\pm$ 9.0\% for Br of $D^\pm\tau\nu$ (Belle II)
  \end{minipage}
 &\begin{minipage}{4.6cm} \small In the year 2012, Babar announced $~3 \sigma$ deviation from the SM. New physics example: charged Higgs.  \end{minipage}\\ \hline
 $Br(B\to X_s\gamma)$ & $(3.15\pm0.23)\times 10^{-4}$& $\pm$ 6 \% (Belle II)& \begin{minipage}{4.6cm} \small Currently SM prediction (at NNLO) is consistent to the experimental value. The error is becoming dominated by the theoretical ones. The result  also depends on $V_{ts}$. New physics example: $b\to s\gamma$ penguin loop\end{minipage}\\ \hline
 $Br(B_{d,s}\to \mu^+\mu^-)$ &
 \begin{minipage}{3cm}$(<1.0  \times 10^{-9})_{B_d} $\\$((3.2^{+1.5}_{-1.2}) \times 10^{-9})_{B_s}$ \end{minipage}
 & \begin{minipage}{3.7cm}$(\pm 0.5(0.15)\times 10^{-9})_{B_s}$\\
  (LHCb)\end{minipage}& \begin{minipage}{4.6cm} \small The result is so far consistent to the SM prediction though it depends on the $|V_{td, ts}|$. The Minimal Flavour Violation (MFV) hypothesis leads a relation $\frac{Br_{B_s\to\mu^+\mu^-}}{Br_{B_d\to\mu^+\mu^-}}=\frac{\hat{B}_d}{\hat{B}_s}\frac{\tau_{B_s}}{\tau_{B_d}}\frac{\Delta M_s}{\Delta M_d}\simeq 30$. Nevertheless, a large enhancement in $B_d\to \mu^+\mu^-$ is possible in non-MFV. New physics example:  large $\tan\beta$ scenarios  \end{minipage}\\ \hline
 $\phi_3=\gamma$ &
 \begin{minipage}{3cm}
 $(69^{+17}_{-16})^{\circ}$ (Babar)\\
 $(68^{+15}_{-14})^{\circ}$ (Belle)\\
 $(71.1^{+16.6}_{-15.7})^{\circ}$ (LHCb)
 \end{minipage}
  &
  \begin{minipage}{3cm}
 $\pm 1.5^{\circ}$ (Belle II)\\
 $\pm 4^{\circ}(0.9^{\circ})$ (LHCb)
 \end{minipage}
  & 
 \begin{minipage}{4.6cm}
\small  The angle $\phi_3(=\gamma)$ is extracted from the decay modes, $DK, D^*K, DK^*$ etc. These are tree level decays that it can be less affected by the new physics  contributions. Therefore, the $\phi_3(=\gamma)$ measurement can be used together with the other purely tree decays, to determine the ``SM'' unitarity triangle to test the new physics contributions to the other box/penguin loop dominant modes. The precision which can be reached in the future is quite impressive and it will be one of the most important measurements in order to fully understand the unitarity triangle.  \end{minipage} \\ \hline
 \end{tabular}
\end{center}
\label{table:strat4}
\end{table}%

\newpage
\section*{Acknowledgements}
We would like to thank to the organizers for giving me the opportunity to give a lecture at AEPSHEP 2012. This work was supported in part by the ANR contract ``LFV-CPV-LHC'' ANR-NT09-508531 and France-Japan corporation of IN2P3/CNRS TYL-LIA.


\begin{thebibliography}{99}
\bibitem{BS}
I.~I.~Bigi and A.~I.~Sanda,
Cambridge Monogr.\ Part.\ Phys.\ Nucl.\ Phys.\ Cosmol.\  {\bf 9} (2000) 1.
\bibitem{PI2}  
  S.~Weinberg,
  ``The Quantum theory of fields. Vol. 1: Foundations,''
  Cambridge, UK: Univ. Pr. (1995) 609 p 
\bibitem{PIII-1} 
  S.~P.~Martin,
  ``A Supersymmetry primer,''
  In *Kane, G.L. (ed.): Perspectives on supersymmetry II* 1-153
  [hep-ph/9709356].
\bibitem{BPBF} 
Book of Physics of B Factories: Babar and Belle collaborations, in preparation
\bibitem{BELLEIITDR}
The Belle II Collaboration, ``Belle II Technical Design Report'', KEK Report 2010-1 [arXiv:1011.0352]
\bibitem{LHCBTDR}
The LHCb Collaboration, ``LHCb reoptimized detector design and performance : Technical Design Report'', CERN-LHCC-2003-030, LHCb-TDR-9


\end{thebibliography}
\end{document}